\documentclass[12pt,pre,aps,showpacs,preprint]{revtex4}

\usepackage{graphics}
\usepackage{graphicx}
\usepackage{amsfonts}
\usepackage{textcomp}
\usepackage{amssymb}
\usepackage{amsmath}

\begin{document}

\title{Geometrical Ambiguity of Pair Statistics. I. Point Configurations}

\author{Y. Jiao}


\affiliation{Department of Mechanical and Aerospace Engineering, 
Princeton University, Princeton New Jersey 08544, USA}

\author{F. H. Stillinger}


\affiliation{Department of Chemistry, Princeton University,
Princeton New Jersey 08544, USA}

\author{S. Torquato}

\email{torquato@electron.princeton.edu}

\affiliation{Department of Chemistry, Princeton University,
Princeton New Jersey 08544, USA}

\affiliation{Princeton Institute for the Science and Technology of
Materials, Princeton University, Princeton New Jersey 08544, USA}

\affiliation{Program in Applied and Computational Mathematics, 
Princeton University, Princeton New Jersey 08544, USA}

\affiliation{Princeton Center for Theoretical Physics, Princeton
University, Princeton New Jersey 08544, USA}

\date{\today}

\pacs{05.20.-y, 61.43.-j}

\begin{abstract}
Point configurations have been widely used as model systems in 
condensed matter physics, materials science and biology. 
Statistical descriptors such as the $n$-body distribution function $g_n$ 
is usually employed to characterize the point configurations, among 
which the most extensively used is the pair distribution function $g_2$. 
An intriguing inverse problem of practical importance that has been receiving considerable 
attention is the degree to which a point configuration can be reconstructed 
from the pair distribution function of a target configuration. 
Although it is known that the pair-distance information contained 
in $g_2$ is in general insufficient to uniquely determine a point configuration, 
this concept does not seem to be widely appreciated 
and general claims of uniqueness of the reconstructions using pair information 
have been made based on numerical studies. In this paper, we 
introduce the idea of the distance space, called the $\mathbb{D}$ space. 
The pair distances of a specific 
point configuration are then represented by a single point in the $\mathbb{D}$ space. 
We derive the conditions on the pair distances that can be associated with 
a point configuration, which are equivalent to the realizability conditions of 
the pair distribution function $g_2$. Moreover, we derive the conditions 
on the pair distances that can be assembled into distinct configurations, 
i.e, with structural degeneracy. 
These conditions define a bounded region in the $\mathbb{D}$ space.  
By explicitly constructing a variety of degenerate point configurations using the $\mathbb{D}$ space, 
we show that pair information is indeed insufficient to uniquely determine the configuration 
in general. We also discuss several important problems in statistical physics 
based on the $\mathbb{D}$ space, 
including 
the reconstruction of atomic structures from experimentally obtained $g_2$ 
and a recently proposed ``decorrelation'' principle.  The degenerate 
configurations have relevance to open questions involving the famous traveling 
salesman problem. 
\end{abstract}

\maketitle

\section{Introduction}

A collection of a finite or infinite number of points in $d$-dimensional 
Euclidean space $\mathbb{R}^d$ is called a \textit{point configuration}. 
Point configurations are one of the most popular and widely used models for many-particle 
systems in various branches of modern science, including 
condensed matter physics and materials science \cite{chaikin, torquato, Sa03, Ed94, zohdi}, 
statistical mechanics \cite{Ri66,weeks, Ha06}, 
discrete mathematics (packing problems) \cite{conway}, 
astrophysics (distribution of galaxy clusters) \cite{Pe93, Ga05}, 
ecology (tree distributions in forests) \cite{ecology} and 
biology (various cellular structures) \cite{Ge08}. Point configurations 
can exhibit a variety of degrees of disorder, from the most random 
Poisson distribution \cite{torquato} to a perfectly ordered Bravais lattice \cite{chaikin}. 
The degrees of disorder can be quantified by discriminating order metrics \cite{chaikin, Tr00},
which, in their simplest forms, are scalars and normalized such that the most disordered 
system is associated with zero and the most ordered ones with unity. 

In most circumstance, it is impossible and even unnecessary to acquire 
detailed knowledge of all positions of the points in the configuration. Instead, 
statistical descriptors such as distribution functions 
are typically employed to characterize the point configurations. 
In particular, the $n$-body distribution function 
$g_n({\bf x}_1,{\bf x}_2,\dots,{\bf x}_n)$ is related to the probability 
of finding a generic configuration of $n$ points at positions ${\bf x}_1,{\bf x}_2,\dots,{\bf x}_n$. 
It is well known that a set of $n$-body distribution functions 
$g_1, g_2, \ldots, g_n$ \cite{torquato} is required to 
statistically characterize an $n$-point configuration completely. As 
$n \rightarrow \infty$ in the thermodynamic limit (e.g., the volume $V$ 
which the $n$ points occupy also increases to infinity such that the number 
of points per volume -- number density $\rho = N/V$ -- is a well defined 
finite number), the set contains an infinite number of correlation functions. 
For statistically homogeneous systems which is the focus 
of this paper, $g_n$ is translationally 
invariant and hence depends only on the relative displacements of the 
positions with respect to some chosen origin, say ${\bf x}_1$, i.e., 
$g_n({\bf x}_1,{\bf x}_2,\dots,{\bf x}_n) = g_n({\bf x}_{12},{\bf x}_{13},\dots,{\bf x}_{1n})$ 
with ${\bf x}_{ij} = {\bf x}_j-{\bf x}_i$. Thus, the one-body distribution 
function $g_1$ is just equal to the number density $\rho$. The important 
two-body quantity $g_2({\bf x}_{12})$ is usually referred to as the pair 
distribution function. In the statistically isotropic case, 
$g_2$ is a radial function, i.e., $g_2({\bf x}_{12}) = g_2(|{\bf x}_{12}|)$ and 
it is also called the radial distribution function. 
The radial distribution function which is one of the most widely used structural descriptors, 
essentially provides the distribution of the 
point-pair separation distances and can be obtained experimentally 
via scattering of radiation \cite{chaikin}. The three-body function 
$g_3$ contains information about how the pair separations involved in 
$g_2$ are linked into triangles. 

It is worth noting that by decorating the points in the system (e.g., 
letting equal-sized spheres be centered at each point), one can construct 
a two-phase random texture from a given point configuration. In general, there 
is an infinite number of ways to decorate a point configuration. 
In the characterization of random textures, the analog of the 
$n$-body distribution functions are the $n$-point correlation functions 
$S_n({\bf x}_1,{\bf x}_2,\dots,{\bf x}_n)$ \cite{torquato}, which gives the probability 
of finding $n$ points at positions ${\bf x}_1,{\bf x}_2,\dots,{\bf x}_n$ in the phase of interest.
In general, a complete statistical characterization of a continuum random texture 
requires an infinite set of $S_n$. Though under certain conditions, 
$g_n$ of a point configuration 
and $S_n$ of the associated decorated random texture might convey 
the same level of structural information 
(in fact the associated $S_n$ can expressed as functional of $g_n$ given the 
details of the decorating phase \cite{torquato}), 
the former evidently reflect the essential geometrical features of the 
point configuration more directly.

An intriguing inverse problem that has been receiving considerable 
attention is the reconstruction (or construction) of realizations of a 
many-body system (essentially a point configuration) 
that match the prescribed structural information of 
the system in the form of $g_2$ or $S_2$, obtained from either experiments 
or theoretical considerations. Examples include the reconstruction of 
random media \cite{Ye98a, Ye98b, Cu99, Utz02, 
ApplyA, ApplyD, Ji07, Ji08, Ka08} and colloidal suspensions \cite{Ri97}, 
investigation of the iso-$g_2$ process \cite{Iso_g2} or $g_2$-invariant 
processes and the realizability conditions of $g_2$ 
\cite{realizable} as well as the more recent 
discovery of  unusual disordered classical 
ground states \cite{Ba08}. X-ray scattering techniques have been 
an indispensable tool historically in the study of the structures of crystalline matter, 
and it has been generalized to probing disordered media \cite{chaikin}. 
In particular, the pair distribution function $g_2({\bf r})$ is obtainable 
from the Fourier transform of the structure factor $S({\bf k})$ \cite{chaikin}, which 
is proportional to the scattering intensity (with the atomic structure function removed) 
and can be directly measured in experiments. 
With the obtained $g_2$, one can then employ 
various reconstruction techniques to generate realizations of
the system of interest. Another related family of inverse problems is the 
reconstruction of pair interaction potential from a given 
radial distribution function $g_2(r)$ between particles, i.e., the 
inverse Monte Carlo problems \cite{Mak07}.

It is known that though the information contained in $g_2$ can be sufficient to 
completely characterize ordered point configurations in very special circumstances \cite{circumstance} 
it is generally devoid of crucial structural 
information to uniquely determine a disordered point configuration \cite{Pa44, Ho54, Bo04}. 
However, it seems that this aspect has not been widely appreciated 
and general claims of uniqueness of the reconstructions using $g_2$ or $S_2$ 
have been made based on numerical studies \cite{Utz02, Ka08}. One aim of this paper is 
to show via a variety of examples the existence of distinct point 
configurations with identical pair-distance distributions (e.g., $g_2$), which 
implies the non-uniqueness of the reconstructions involving $g_2$ of these 
point configurations. Besides, general mathematical formalism to characterize 
the structural ambiguity of pair information is also devised.


\begin{figure}[bthp]
\begin{center}
$\begin{array}{c@{\hspace{1.25cm}}c}\\
\includegraphics[width=5.5cm,keepaspectratio]{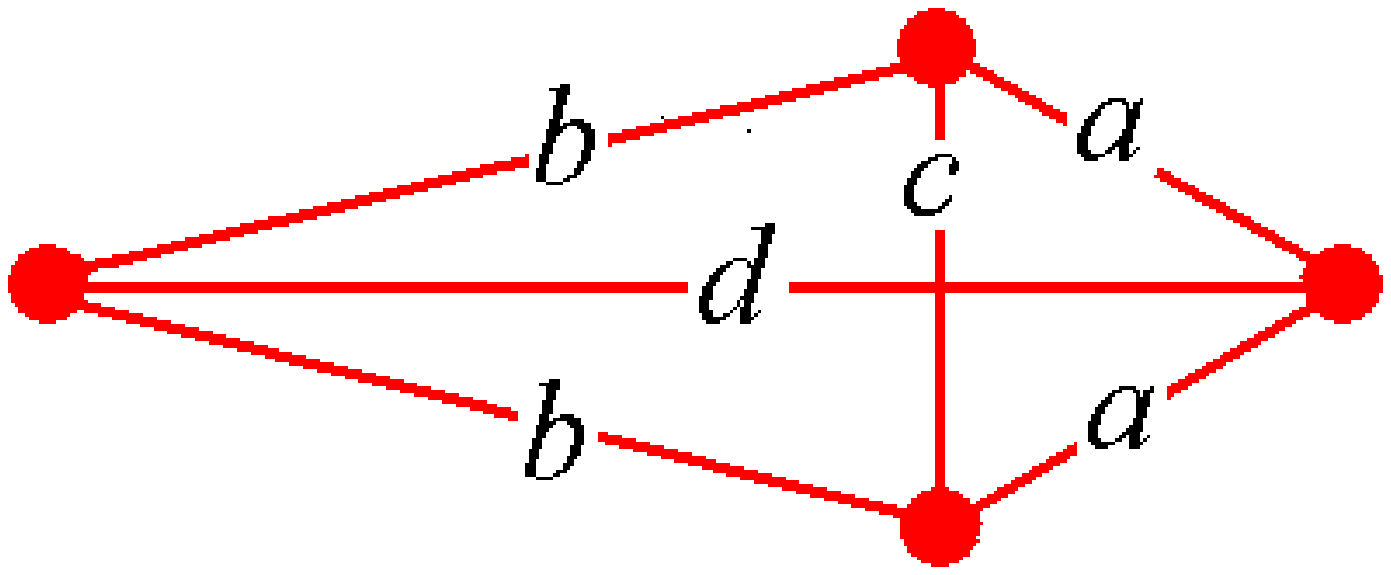} &
\includegraphics[width=5.5cm,keepaspectratio]{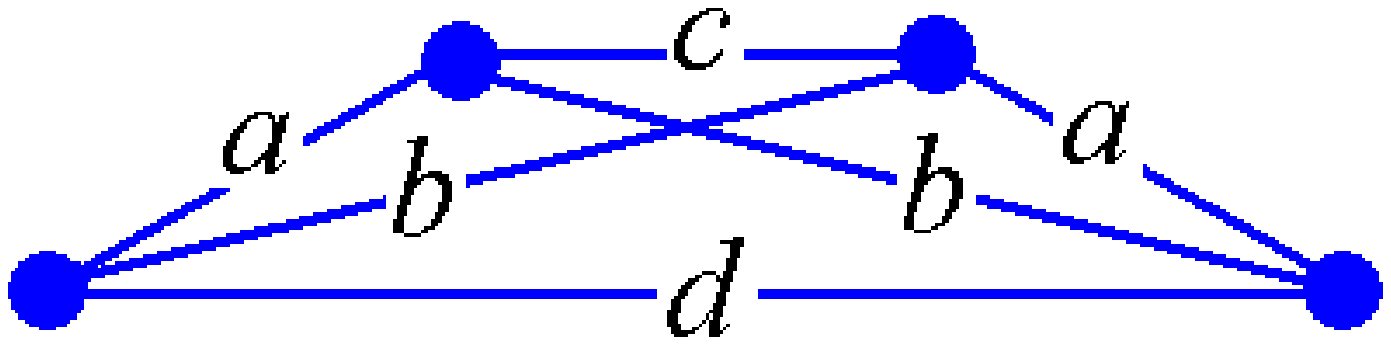} \\
\mbox{\bf (a)} & \mbox{\bf (b)}
\end{array}$
\end{center}
\caption{(color online). An example of two-dimensional four-point configurations possessing 
two-fold degeneracy: (a) A ``kite''. (b) A ``trapezoid''. The specific distance 
sets are $a=(2x^2-3x+\frac{5}{4})^{\frac{1}{2}}$, $b=(2x^2-x+\frac{1}{4})^{\frac{1}{2}}$, 
$c=2x-1$, $d=1$, for $\frac{1}{2}<x<1$. For $x>1$, the outer boundary of the ``kite'' is no 
longer a quadrilateral but reduces to an isosceles triangle.} \label{kite-trape}
\end{figure}

Figure \ref{kite-trape} shows two distinct configurations of four 
points in two dimensions with identical pair distances. 
In particular, one configuration (with the pair distances shown) 
resembles a ``kite'' and the other resembles a ``trapezoid''. 
In order to provide an in-depth presentation of the ambiguity of 
pair-distance distributions, it is necessary to exam the problem 
mathematically first and then discuss the physical implications. 
Some definitions are in order here.
Two $d$-dimensional statistically homogeneous and isotropic 
$n$-point configurations $\Gamma^i_{d,n}$ and $\Gamma^j_{d,n}$ are 
identical if and only if they possess identical sets of 
$k$-body distribution functions $g_k$ for $k=1, 2, \ldots, n$. 
The configurations $\Gamma^i_{d,n}$ and $\Gamma^j_{d,n}$ are $g_k$-distinct if and only if they process 
distinct $n$-body distribution functions for all $n \ge k$.
A $d$-dimensional $n$-point configuration $\Gamma^1_{d,n}$ is $k$-fold degenerate 
if and only if there exist additional $(k-1)$ $d$-dimensional $n$-point configurations 
$\Gamma^i_{d,n}$ ($i=2,\ldots, k$) that are mutually $g_3$-distinct and also 
$g_3$-distinct from $\Gamma^1_{d,n}$, all of which possess the same two-body distribution 
function $g_2$. This definition of structural degeneracy rules out the possibility 
that two degenerate point configurations are trivially connected by 
translation, rotation, mirror reflection or any of their combinations. Moreover, 
we consider that two point configurations are equivalent (i.e., do not 
form a degenerate pair) if they are related by 
a trivial isotropic rescaling, which does not change the internal structure of the configuration. 
Thus, we see that the ``kite'' and the ``trapezoid'' are associated with the 
same set of pair distances (i.e., they are two-fold degenerate), but the triangle information 
of the two is distinguishable \cite{Frank00}. It is worth noting 
the historically prominent Kirkwood superposition approximation of $g_3$ 
which replaces the three-body distribution function with a product of three pair distribution functions \cite{Kirkwood}.  
Because the separate members of our pair-distance degeneracy examples present distinct triangle 
(i.e. three-body) distributions, the conclusion must be that no functional of $g_2$ 
(Kirkwood or otherwise) can uniquely specify $g_3$.

It is clear that given $g_2$ associated with the degenerate point configurations, 
it is impossible even in principle to obtain a unique reconstruction, 
and each degenerate configuration should be recovered with equal probability. 
Therefore, an outstanding problem is to determine under what conditions 
the pair distance information contained in $g_2$ could uniquely 
determine a point configuration, i.e., there is no associated 
structural degeneracy. A question with more practical importance 
is that how the point configurations would change when the measurement of 
$g_2$ is subject to slight imprecision, 
a common situation in experiments and numerical simulations.

\begin{figure}[bthp]
\begin{center}
$\begin{array}{c}
\includegraphics[width=10.5cm,keepaspectratio]{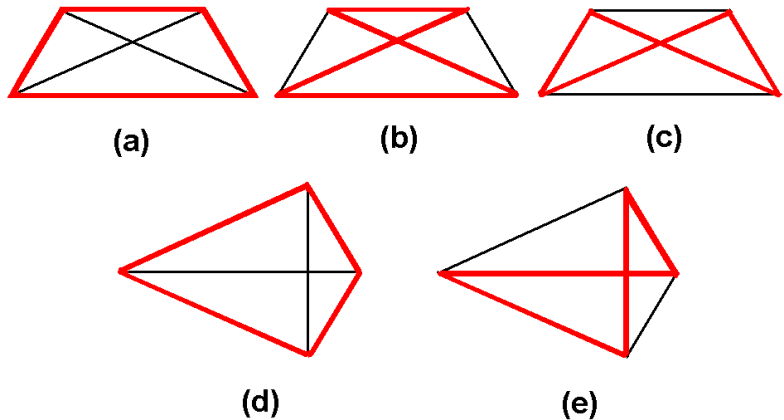}
\end{array}$
\end{center}
\caption{(color online). The three distinguishable circuits for the ``trapezoid'' (upper panel) 
and two distinguishable circuits for the ``kite'' (lower panel). The circuits are 
shown in thick red lines. The circuit shown in (a) is the shortest route among all possibilities.} \label{TSP_fig}
\end{figure}

In addition to their physical relevance, degenerate point 
configurations are also of mathematical interest. For example, 
open questions connected to the famous traveling salesman problem \cite{STP} 
can be raised: Given the degenerate configurations associated with 
the same set of pair distances, what are the optimal solutions of 
the traveling salesman problem  for each configuration and are they unique? 
Are there special degenerate configurations whose solutions are identical?
For the simplest ``kite-trapezoid'' example shown in Fig.~\ref{kite-trape}, 
for the parameter values $x>1/2$ the ``trapezoid'' has three distinguishable circuits 
and the ``kite'' has only two (see Fig.~\ref{TSP_fig}). The shortest route among all 
is presented by the ``trapezoid'', i.e., a closed circuit visiting 
each vertex once and only once. For $x=1/2$, both the ``kite'' and the ``trapezoid'' 
collapse onto a line segment, and in that limiting case all circuits have the same length. 
For more general and complicated degenerate configurations, 
such questions are notoriously difficult to solve; the problem belongs to the NP-complete class.

\begin{figure}[bthp]
\begin{center}
$\begin{array}{c@{\hspace{1.5cm}}c}\\
\includegraphics[width=4.25cm,keepaspectratio]{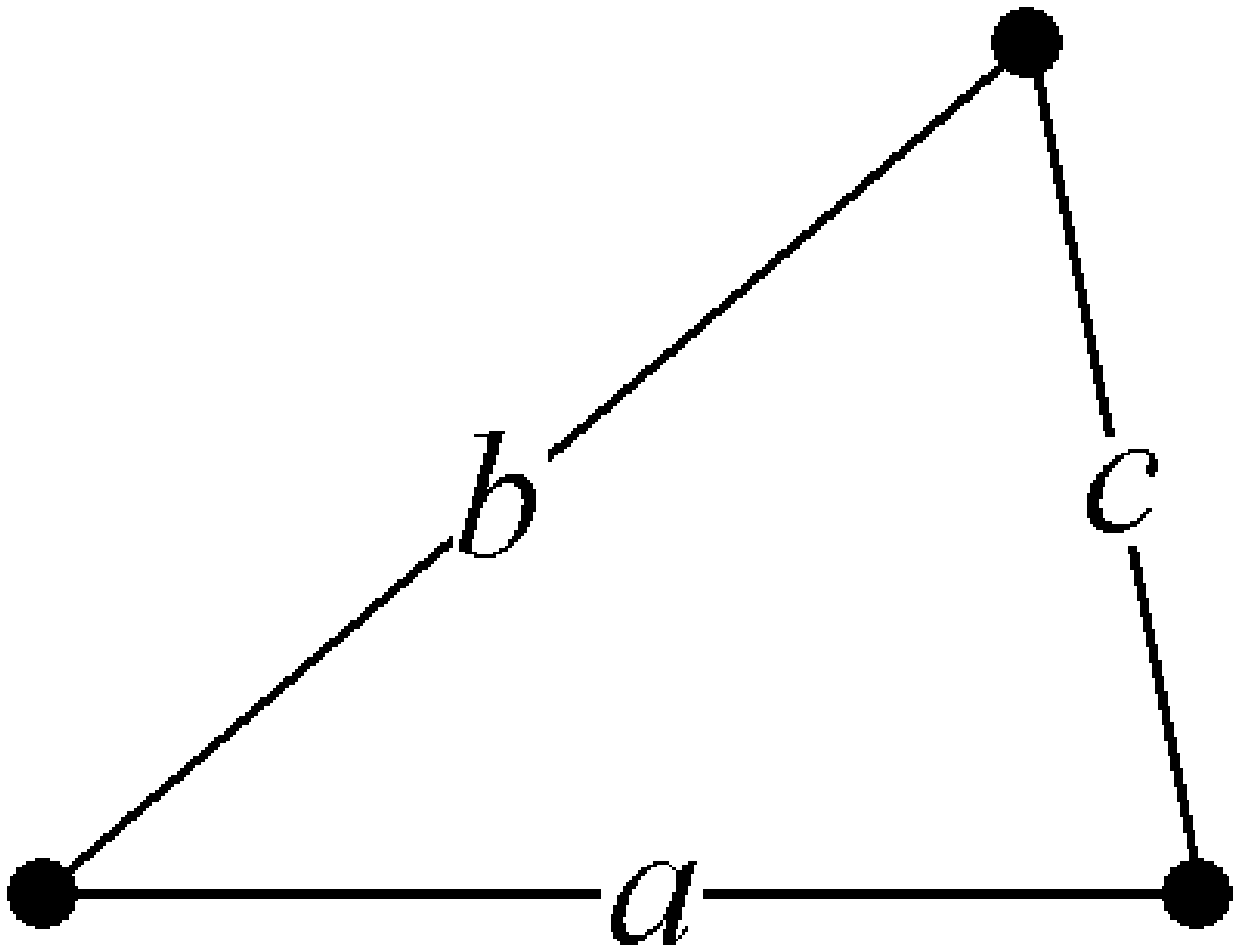} &
\includegraphics[width=4.75cm,keepaspectratio]{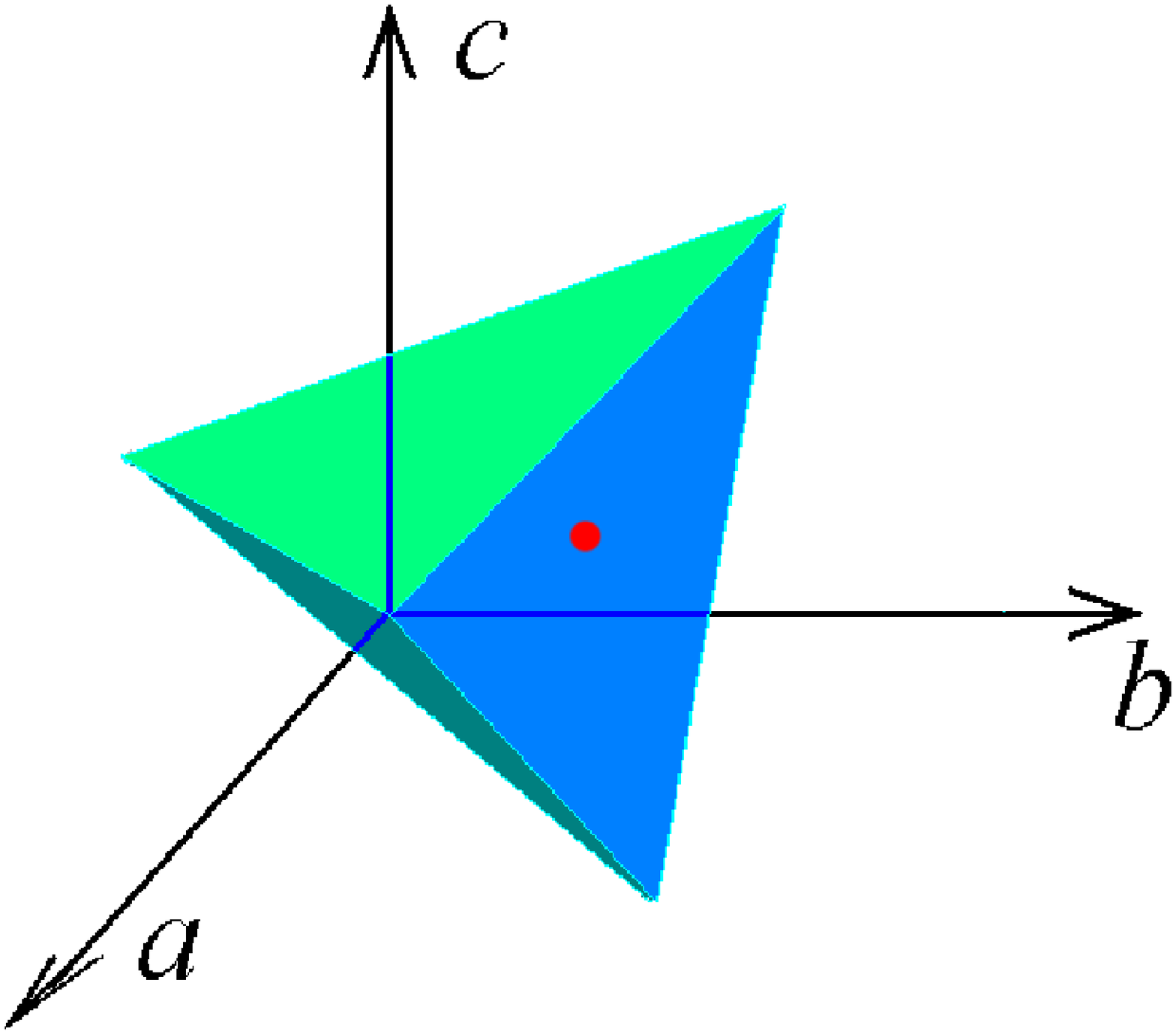} \\
\mbox{\bf (a)} & \mbox{\bf (b)}
\end{array}$
\end{center}
\caption{(color online). (a) A three-point configuration (i.e., a triangle) in $\mathbb{R}^2$. 
(b) The region of feasible distances in the $\mathbb{D}$ space (bounded by the 
blue planes). The three pair distances of the triangle shown in (a) is represented as 
a point (red spot) in the $\mathbb{D}$ space.} \label{2DP3-Dspace}
\end{figure}

In this paper, we introduce the idea of the distance space (i.e., the $\mathbb{D}$ space), 
in which each dimension is associated with the separation distance between a 
given point-pair. 
The pair-distance distribution  
of a particular point configuration is then presented by a single point in the $\mathbb{D}$ space. 
It is clear that not all the points in $\mathbb{D}$ space correspond to 
realizable configurations, i.e., the separation distances have to satisfy 
certain conditions such that they could be assembled into a point configuration. 
These conditions together define a (partially) bounded region in the 
$\mathbb{D}$ space. For example, for three-point configurations in $\mathbb{R}^2$ (i.e., triangles), 
the region of the feasible distances is an open ``pyramid'' in the three-dimensional 
$\mathbb{D}$ space, as shown in Fig.~\ref{2DP3-Dspace}. When degeneracy exists the 
region of the feasible distances is generally a complicated closed intersection of several 
such simple curved ``pyramid'' in high dimensions. 
The determination of the region of feasible distances is equivalent 
to obtaining the conditions of a realizable $g_2$, i.e., a pair distribution 
function that can be associated with a point configuration. 
Using the $\mathbb{D}$ space, 
we will answer various aspects of the aforementioned questions concerning the 
degenerate point configurations and the non-uniqueness issue of the reconstruction. 
We will show that the utility of the $\mathbb{D}$ space also improves 
our understanding of various important 
problems in statistical physics such as the recently proposed decorrelation principle 
in high-dimensional Euclidean space \cite{To06}. 
In a sequel to this paper, we will extend the present analysis  
to understand  degeneracy issues pertaining to heterogeneous materials,
which is a larger classification than point configurations \cite{Ji09b}.

The rest of the paper is organized as follows: In Sec.~II, we discuss the 
$\mathbb{D}$ space in detail and derive the conditions for feasible distances 
and for the occurrence of degeneracy, through which we show that in general 
degeneracy is rare. In Sec.~III, we provide a variety of examples of degenerate 
point configurations and illustrate how the conditions derived in Sec.~II 
could be employed to construct point configurations with specific degeneracy. 
In Sec.~IV, we discuss several problems in statistical physics such as the 
reconstruction 
of atomic structures from experimentally obtained $g_2$ and the decorrelation principle, 
based on the idea of the $\mathbb{D}$ space. Finally, we make concluding remarks.

\section{The Distance Space $\mathbb{D}$}

In this Section, we will discuss in details the $\mathbb{D}$ space. In particular, 
we will derive the conditions under which the pair distances could be 
assembled into a point configuration, i.e., the feasibility conditions 
as well as the conditions under which the pair distances correspond to 
degenerate point configurations. 
We will first study a four-point configuration in $\mathbb{R}^2$ to 
illustrate the idea and then consider the general $n$-point configurations 
in $\mathbb{R}^d$.

\subsection{A Simple Example: Four-Point Configuration in $\mathbb{R}^2$}

\begin{figure}[bthp]
\begin{center}
$\begin{array}{c}
\includegraphics[width=6.25cm,keepaspectratio]{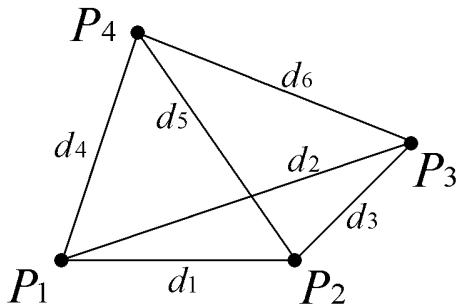}
\end{array}$
\end{center}
\caption{A four-point configuration in $\mathbb{R}^2$.} \label{2D4P}
\end{figure}

Consider a four-point configuration $\Gamma_{2,4}$ in $\mathbb{R}^2$ 
(see Fig.~\ref{2D4P}) and the associated 
6-dimensional $\mathbb{D}$ space. We would like to know 
the answers to the following two questions:

\noindent{(i)} {\sl What are the conditions 
the six pair separation distances must satisfy so that they correspond 
to a four-point configuration in $\mathbb{R}^2$? }

\noindent{(ii)} {\sl What are the conditions 
the pair distances must satisfy so that they correspond to $k$-fold 
degenerate four-point configurations in $\mathbb{R}^2$?}

To answer these questions, we consider a particular construction as follows: Suppose 
the six pair distances are elements of the set $\Omega = \{d_1, d_2, \ldots, d_6\}$, 
which can be further partitioned as $\Omega = \{P_1, P_2, P_3, P_4\}$ where 
$P_1 = \{\Phi\}$ ($\Phi$ is the null set), $P_2 = \{d_1\}$, $P_3 = \{d_2, d_3\}$ and 
$P_4 = \{d_4, d_5, d_6\}$. We will see that such a partition enables us to associate the 
pair distances with the corresponding points in a convenient way. 
Recall that from our definition (Sec. I), point configurations 
are considered identical if they are connected by translation, rotation, 
mirror reflection and any of their combinations. Thus, we can put point $P_1$ 
at the origin of a Cartesian coordinate system and put point $P_2$ on one of 
the two orthogonal axes of the coordinate system separated from the origin (i.e., $P_1$) 
by a distance $d_1$. Note different choices of the position of $P_1$ and the 
orientation of the line segment $\overline{P_1P_2}$ lead to point configurations that 
are identical up to translations and rotations. 
For point $P_3$, we can either let $\overline{P_1P_3} = d_2$, 
$\overline{P_2P_3} = d_3$ or $\overline{P_1P_3} = d_3$, $\overline{P_1P_2} = d_2$. 
The two choices correspond to two configurations connected by a mirror 
reflection, which are considered identical and either choice is acceptable. 
With out loss of generality, we choose $\overline{P_1P_3} = d_2$, 
$\overline{P_2P_3} = d_3$. Finally, we choose $\overline{P_1P_4} = d_4$, 
$\overline{P_2P_4} = d_5$, $\overline{P_3P_4} = d_6$ for point $P_4$.

We see from the above construction that the positions of points $P_3$ and $P_4$ 
are determined with respect to the line segment defined by points $P_1$ and $P_2$ 
as a ``reference'' structure. Note that the line segment is a one-dimensional simplex. 
In $\mathbb{R}^2$, the position of a point is completely determined by specifying 
two distances from the point of interest to the vertices of a reference line segment, 
given that the distances involved satisfy the triangular inequality, i.e., the 
triangle formed by the point of interest and the two vertices of the reference 
line segment possess non-negative area. The area $\Delta$ of a triangle with edges $a$, $b$, $c$ 
is related to the Cayley-Menger determinant \cite{So58}, i.e.,

\begin{equation}
\label{eq01}
\Delta^2 = -\frac{1}{16}\left|{\begin{array}{c@{\hspace{0.3cm}}c@{\hspace{0.3cm}}c@{\hspace{0.3cm}}c} 
0 & 1 & 1 & 1 \\ 
1 & 0 & a^2 & b^2 \\ 
1 & a^2 & 0 & c^2 \\ 
1 & b^2 & c^2 & 0 \\\end{array}}\right|.
\end{equation} 

\noindent Thus, for point $P_3$, we obtain 

\begin{equation}
\label{eq02}
d_1^4+d_2^4+d_3^4-2d_1^2d_2^2-2d_1^2d_3^2-2d_2^2d_3^2 > 0,
\end{equation}

\noindent and for point $P_4$, we obtain 

\begin{equation}
\label{eq03}
d_1^4+d_4^4+d_5^4-2d_1^2d_4^2-2d_1^2d_5^2-2d_4^2d_5^2 > 0.
\end{equation}

\noindent Inequalities (\ref{eq02}) and (\ref{eq03}) define a partially bounded region in the 
six-dimensional $\mathbb{D}$ space, the lower-dimensional analog of which 
is the open pyramid shown in Fig.~{\ref{2DP3-Dspace}}. The distance $d_6$ between 
points $P_3$ and $P_4$ is also completely determined by $d_1, d_2, \ldots, d_5$ via

\begin{equation}
\label{eq05}
\left|{\begin{array}{c@{\hspace{0.35cm}}c@{\hspace{0.35cm}}c} 
-2d_1^2 & (d_3^2-d_1^2-d_2^2) & (d_5^2-d_1^2-d_4^2)  \\ 
(d_3^2-d_1^2-d_2^2) & -2d_2^2 & (d_6^2-d_2^2-d_4^2)  \\ 
(d_5^2-d_1^2-d_4^2) & (d_6^2-d_2^2-d_4^2) & -2d_5^2 \\\end{array}}\right| = 0,
\end{equation}

\noindent which results from the requirement that all the $3\times3$ 
minors of the Gram matrix \cite{Bo04} involving the distances possess zero determinant.
We will discuss the Gram matrix in detail in Sec.~II.B. Equation(\ref{eq05}) defines a 
curved hypersurface in the $\mathbb{D}$ space, whose intersection with the 
region defined by Eqs.~(\ref{eq02}) and (\ref{eq03}) contains the feasible 
distances $\Omega$ that can be assembled into a four-point configuration in $\mathbb{R}^2$. 
We call Eqs.~(\ref{eq02}), (\ref{eq03}) and (\ref{eq05}) the \textit{feasibility} conditions. 
Note that for the four-point configuration in $\mathbb{R}^2$, only five 
pair distances can be chosen almost independently subject to the mild 
triangle inequality constraint. Thus we define the \textit{free dimension} 
of the $\mathbb{D}$ space to be the number of pair distances that 
are only constrained by inequalities, which total to five here. 
The free dimension is also the dimension of the region for the feasible distances, 
which is also referred to as the \textit{feasible region}.

Now we can answer question (i) given at the beginning of this Section 
easily. Suppose a list of distances is given, when any one permutation 
of these distances satisfies the feasibility conditions [Eqs.~(\ref{eq02}),(\ref{eq03}) 
and (\ref{eq05})], the pair distances correspond to a four-point configuration in $\mathbb{R}^2$. 
However, such a simple answer does not exist for (ii). For the pair distances 
to correspond to $k$-fold degenerate point configurations, a necessary condition is 
that the dimension of the intersection of the feasible regions for the $k$ permutations of the 
pair distances is non-zero. In other words, each permutation of the pair distances is 
associated with a set of feasibility conditions, 
and a feasible region can be constructed. To obtain a $k$-fold degeneracy, 
all sets of the feasibility conditions need to be satisfied simultaneously, which is only 
possible when the intersection of the feasible regions is at least 
a single curve in the $\mathbb{D}$ space. For the two-dimensional four-point configuration 
of interest the free dimension is five, 
which leads to an upper bound on the order of the degeneracy, i.e., $k_{max} = 5$. 
This condition is only a necessary one because there are certain 
permutations that lead to identical configurations, such as those that correspond 
to the permutations among the point indices which do not change 
the structure of the configuration, since the points are indistinguishable. For example, 
$\Omega_1 = \{d_1, d_2, d_3, d_4, d_5, d_6\}$ and $\Omega_2 = \{d_1, d_4, d_5, d_2, d_3, d_6\}$ 
correspond to an exchange of $P_3$ and $P_4$, 
which possess the identical feasible regions and thus do not contribute the 
the degeneracy. No further conclusions can be made without knowing the details 
of how the distances are permuted. In Sec.~III
we will construct concrete examples of degenerate $\Gamma_{2,4}$, 
where the details of permutations are considered.

\subsection{General Formulation: Feasibility Conditions}


The generalization of the above formulation is straightforward. Note in 
the following discussion in this Section, we assume $n>(d+1)$; the case 
when $n=d+1$ (i.e., the simplex configurations) are discussed in detail 
in Sec.~III.A and the case when $n<(d+1)$ is trivial. Consider 
an $n$-point configuration $\Gamma_{d,n}$ in $\mathbb{R}^d$, which 
possesses $m=n(n-1)/2$ pair distances $\Omega = \{d_1, d_2, \ldots, d_m\}$. 
The distances can be further partitioned, i.e., $\Omega = \{P_1, P_2, \ldots, P_n\}$, 
where $P_1 = \{\Phi\}$, $P_2 = \{d_1\}$, $\ldots$, 
$P_i = \{d_{(i^2-3i+3)/2},\ldots, d_{(i^2-i)/2}\}$, $\ldots$, 
$P_n = \{d_{(n^2-3n+3)/2},\ldots, d_{(n^2-n)/2}\}$. 
Following the same construction procedure prescribed in Sec.~II.A, the distances 
associated with the first $d$ points, i.e., $P_1, P_2, \ldots, P_d$, are
assembled into a $(d-1)$-dimensional simplex as the ``reference'' structure. 
Each point $P_i$ ($i>d$) is associated with $(i-1)$ distances and the 
position of point $P_i$ is completely determined by specifying the $d$ 
distances from $P_i$ to the vertices of the reference structure, 
given that the $d$-dimensional simplex formed by $P_i$ and the vertices of the 
reference structure possesses a nonnegative volume. In particular, 
denote the $(d+1)$ vertices of the $d$-dimensional simplex by ${\bf v}_i$ ($i=1,\ldots, d+1$), 
we can define a $(d+1)\times(d+1)$ distance matrix $M$, i.e., 

\begin{equation}
\label{eq06}
M_{ij} = M_{ji} = || {\bf v}_i - {\bf v}_j ||^2,
\end{equation}

\noindent where $||\cdot||$ denotes the $L^2$-norm of a $d$-dimensional vector 
and $M_{ij}$ ($M_{ji}$) is the 
squared distance between vertex $i$ and $j$. The volume $\Delta$ of the simplex is then given 
by the Cayley-Menger determinant, i.e., 

\begin{equation}
\label{eq07}
\Delta^2_d = \frac{(-1)^{d+1}}{2^d(d!)^2} |\hat{M}| \ge 0
\end{equation}

\noindent where $\hat{M}$ is a $(d+2)\times(d+2)$ matrix obtained from $M$ 
by bordering $M$ with a top row $(0,1,...,1)$ and a left column $(0,1,...,1)^T$. 
For example, Eq.~(\ref{eq07}) reduces to Eq.~(\ref{eq01}) in $\mathbb{R}^2$, and 
in $\mathbb{R}^3$ we obtain

\begin{equation}
\label{eq08}
\Delta^2 = \frac{1}{288}\left|{\begin{array}{c@{\hspace{0.3cm}}c
@{\hspace{0.3cm}}c@{\hspace{0.3cm}}c@{\hspace{0.3cm}}c} 
0 & 1 & 1 & 1 & 1\\ 
1 & 0 & M_{12} & M_{13} & M_{14} \\ 
1 & M_{21}& 0  & M_{23} & M_{24} \\ 
1 & M_{31} & M_{32} & 0 & M_{34} \\ 
1 & M_{41} & M_{42} & M_{43} & 0  \\\end{array}}\right|.
\end{equation}

\noindent The requirement that the constructed 
$d$-dimensional simplex possesses non-negative 
volume leads to higher dimensional analogs of the well 
known triangle inequalities in two dimensions, which 
we will refer to as \textit{simplex inequalities}. In general, 
each set of the simplex inequalities associated with a 
point $P_i$ ($i>d$) defines a partially bounded region 
in the $\mathbb{D}$ space, the intersection of which 
is a high-dimensional analog of the open pyramid 
shown in Fig.~\ref{2DP3-Dspace}(b).

It is clear from the above construction that a point configuration 
$\Gamma_{d,n}$ can be completely determined by only specifying 
$f=[\frac{1}{2}d(d-1)+(n-d)d]$ distances (e.g., ``free'' distances), satisfying the 
simplex inequalities. Thus, the free dimension of the $\mathbb{D}$ space 
is $f$ and the remaining $(m-f)$ pair distances (e.g., ``constrained'' distances) 
cannot be chosen freely but instead are completely determined by the $f$ 
``free'' distances. To obtain the relations between the ``constrained'' and 
``free'' distances, we will employ the following theorem \cite{Bo04}: 
 
\smallskip
\noindent{Theorem 1:} {\sl For a set of $n$ vectors ${\bf v}_1, {\bf v_2}, \ldots, {\bf v}_n$
in $\mathbb{R}^d$ ($n>d$), let the Gram matrix be defined by $G_{ij} = <{\bf v}_i, {\bf v}_j>$ 
where $<\cdot>$ denotes the inner product. Then all $(d+1)\times(d+1)$ minors of $G$ must have 
zero determinant.}
\smallskip

\noindent The proof of the theorem is given in Ref.~\cite{Bo04}. It is essentially 
another way of stating the fact that there are at most $d$ linearly 
independent vectors among ${\bf v}_1, {\bf v_2}, \ldots, {\bf v}_n$ in a 
$d$-dimensional Euclidean space. Without loss of generality, 
we can choose the origin at ${\bf v}_1$ and obtain
\begin{equation}
\label{eq09}
\hat{G}_{ij} = \langle \hat{\bf v}_i, \hat{\bf v}_j \rangle = \langle{\bf v}_i-{\bf v}_1, {\bf v}_j-{\bf v}_1\rangle.
\end{equation}

\noindent Consider the identity \cite{Bo04},
\begin{equation}
\label{eq10}
\langle{\bf v}_i-{\bf v}_1, {\bf v}_j-{\bf v}_1\rangle = \frac{1}{2}
[\langle{\bf v}_i-{\bf v}_1, {\bf v}_i-{\bf v}_1\rangle+\langle{\bf v}_j-{\bf v}_1, {\bf v}_j-{\bf v}_1\rangle-
\langle{\bf v}_i-{\bf v}_j, {\bf v}_i-{\bf v}_j\rangle],
\end{equation}

\noindent we obtain that
\begin{equation}
\label{eq11}
\hat{G}_{ij} = \frac{1}{2}(\hat{d}_{i1}^2 + \hat{d}_{j1}^2 - \hat{d}_{ij}^2),
\end{equation}

\noindent where $\hat{d}_{ij}$ is the distance between the two points $i$ and $j$. 
Thus we see the requirement that all $(d+1)\times(d+1)$ minors of $\hat{G}$ have zero 
determinant (denoted by $M_{(d+1)}$), i.e., 
\begin{equation}
\label{eq12}
|M_{(d+1)}(\hat{G})| = 0
\end{equation}

\noindent leads to fourth order algebraic equations involving the 
$(m-f)$ ``constrained'' distances. It is clear that each 
``constrained'' distance can be explicitly expressed as 
a function of the ``free'' distances alone. For four-point 
configuration in $\mathbb{R}^2$, Eq.~(\ref{eq12}) gives 
Eq.~(\ref{eq05}). Note these equalities define 
curved hypersurfaces in the $\mathbb{D}$ space. The 
intersection of these curved hypersurfaces as well as 
the partially bounded regions defined by (\ref{eq07}) gives 
the feasible region of the $\mathbb{D}$ space, i.e., 
when any permutation of the $m = n(n-1)/2$ pair distances lies within the 
feasible region, these distances can be assembled into 
an $n$-point configuration in $\mathbb{R}^d$.

\subsection{General Formulation: Necessary Conditions of Degeneracy}


For the distances $\Omega$ to correspond to $k$-fold degenerate 
point configurations $\Gamma_{d,n}^1$, $\Gamma_{d,n}^2$, $\ldots$, $\Gamma_{d,n}^k$ 
($k \le f$), the feasibility conditions for $k$ distinct permutations 
of $\Omega$ should be satisfied simultaneously. The feasibility 
conditions associated with any particular permutation of $\Omega$ include 
a set of equalities, which would reduce the dimension of the feasible 
region in the $\mathbb{D}$ space. Suppose that for each distinct 
permutation only one additional equality constraint is introduced. 
Then we can obtain an upper bound on the order of the degeneracy, i.e., $k_{max} = f$, 
which corresponds to a feasible region that has been reduced 
to a single curve (with one free dimension). That is, only one distance can 
be chosen arbitrarily. However, different choices of the single free 
distance correspond to trivial isotropic rescaling of the entire configuration, 
which leads to no degeneracy based on our definition.
Note that if the 
permutation does not introduce new feasibility conditions, it corresponds 
to a permutation of the point indices, which does not affect 
the structure of the configuration.

The properties of the feasible regions have important implications. 
As we have seen, the structural degeneracy would reduce the dimension 
of the feasible regions, the volume of which is proportional to the 
number of feasible distance sets. For a particular $n$-point 
configuration, we could in principle identify all feasible distance 
sets by exploring the whole feasible region in the $\mathbb{D}$ space point by point. 
However, the distance sets associated with degeneracies can only 
lie on a hypersurface with lower dimensions than the feasible region. 
The volume ratio of the hypersurface to the feasible region, which 
is also the number ratio of the distance sets associated with 
degeneracies to those without degeneracies, is vanishingly small. 
In other words, although degeneracies exist they are extremely rare. 
This might explain why perfect reconstructions (identical match 
of the pair distances and the configurations up to translations, 
rotations and mirror reflections) can be obtained numerically \cite{Utz02, Ka08}. 
However, the general conclusion that pair statistics alone 
would uniquely determine the configurations could not 
be made only based on those numerical results, as we will 
show in the next section via a variety of examples of degeneracy.

\section{Examples of Degenerate Point Configurations}

In this Section, we construct a variety of degenerate point configurations 
using the general scheme developed in Sec.~II. In particular, we study the 
degeneracies of simplices in $\mathbb{R}^d$, four-point configurations in 
$\mathbb{R}^2$ and specific $n$-point configurations in $\mathbb{R}^d$ possessing 
two-fold degeneracy.

\subsection{Degenerate $d$-Dimensional Simplices}

A simplex in $\mathbb{R}^d$ is the convex hull of a set of $(d+1)$ points $\Gamma_{d,d+1}$
that do not all lie on the same $(d-1)$-dimensional hyperplane. A 
simplex in $\mathbb{R}^2$ is a triangle and in $\mathbb{R}^3$ is a tetrahedron. 
Simplices in $\mathbb{R}^d$ ($d\ge4$) can be considered to be $d$-dimensional generalizations of 
the three-dimensional tetrahedron. The simplex is so-named because it 
represents the simplest possible polyhedron in the given dimension. 
The volume a $d$-dimensional simplex is given by Cayley-Menger determinant (\ref{eq07}).

A unique feature of simplex configurations $\Gamma_{d,d+1}$ is 
that their feasibility conditions only include the 
simplex inequalities. These inequalities define a partially 
bounded region possessing the same dimensions as the $\mathbb{D}$ space. 
In other words, the free dimension of the feasible region is 
not reduced due to degeneracy. Thus, one should expect that 
it is much easier to obtain highly degenerate simplices than other point configurations.

Suppose we have a distance set $\Omega = \{d_1, d_2, \ldots, d_m\}$ ($m = d(d+1)/2$). 
It is clear that if we choose $d_i = \overline{d} + \delta_i$, where $\delta_i$ 
($i=1, 2, \ldots, m$) are mutually distinct small numbers.
they will satisfy all the simplex inequalities and correspond to a point 
in the vicinity of the centroid of the feasible region. The maximum magnitude of 
the $\delta$'s depends on the boundaries of the feasible region, which we need 
not to worry about for the moment, as long as the $\delta$'s are sufficiently 
small and mutually distinct. 

In $\mathbb{R}^2$ ($d=2$), the three distances can be assembled into 
a unique triangle, i.e., we have $k_{max}^{(2)} = 1$. This can also be 
seen from the following argument: since configurations connected by 
translations and rotations are considered identical, we could pick any 
one of the three distances along one of the coordinate axes starting 
from the origin and require the same for the corresponding distance of 
all possible degenerate configurations. In this way, we rule out 
translations and rotations in a plane. There are only two distances left, which 
can be assembled into a triangle in two ways. However, the two 
resulting triangles are mirror image of each other. Thus, we have 
$k_{max}^{(2)} = {2!}/{2} = 1$, where ``!'' indicates factorial.

In $\mathbb{R}^3$ ($d=3$), we similarly choose one of the six 
distances as the ``reference'' distance, and the remaining five 
are assigned to different edges of a tetrahedron, which results 
in $5!$ tetrahedra. However, among these tetrahedra there are 
pairs that are connected by mirror reflection which has to be 
excluded. Two mirror reflection plane can be identified: one 
perpendicular to the reference distance and the other contains 
the reference distance. This further reduces the number of distinct 
tetrahedra by a factor of $1/4$. Thus, we obtain $k_{max}^{(3)} = {5!}/{(2\times2)} = 30$.

Generally, in $d$-dimensions when one of the $m=d(d+1)/2$ distances 
is chosen as the reference distance, there are $(m-1)!$ ways to
assemble the remaining $(m-1)$ distances into a simplex in $\mathbb{R}^d$. 
However, $(d-1)$ hyperplanes (among which one contains the reference 
distance and the others are perpendicular to it) can be identified 
that are mirror reflection hyperplanes of the simplex. Each mirror 
reflection reduces the number of distinct simplices by a factor of $1/2$. 
Thus, we have 

\begin{equation}
\label{eq13}
k_{max}^{(d)} = \frac{(m-1)!}{2^{(d-1)}} = \frac{[d(d+1)/2-1]!}{2^{(d-1)}}.
\end{equation}

We can see that for simplex configurations in $\mathbb{R}^d$ with $d\ge 3$, 
$k_{max}^{(d)}$ is significantly larger than the dimension of the feasible 
region $f = d(d+1)/2$, which indeed implies a high level of degeneracies 
associated with these configurations. 

\subsection{Two-Dimensional Four-Point Configurations}

We show here how the conditions determining the feasible region in $\mathbb{D}$ space 
can be employed to construct four-point configurations $\Gamma_{2,4}$ in $\mathbb{R}^2$ 
with $k$-fold degeneracy. As pointed out in Sec.~II, the feasibility 
conditions are only necessary and the details of how the distances are permuted 
must be considered.

The relations of the six distances of $\Gamma_{2,4}$ 
are given by Eq.~(\ref{eq05}) for the particular 
order $\Omega = \{d_1, d_2, d_3, d_4, d_5, d_6\}$. For a permutation $\Omega^*$, the 
variable $d_i$ in Eq.~(\ref{eq05}) should be replaced by the $i$th element of $\Omega^*$, 
which generally would lead to a different equation for the six distances. 
As mentioned in the last Section, we could choose $d_1$ as the reference distance 
to rule out translation and rotation and only consider the permutations of the 
remaining five distances, which gives $5! = 120$ distinct equations. Without loss of 
generality, we could also choose $d_1 = 1$ which corresponds to a trivial isotropic 
rescaling of the entire point configuration.

In principle, a $k$-fold degeneracy ($k\le k_{max} = 5$) could be constructed by requiring that 
the $k$ equations for the five distances corresponding to $k$ distinct permutations 
hold simultaneously. However, we find that high level degeneracies (those with $k$ close 
to $k_{max}$) are difficult to realize. In particular, when $k$ is large the 
equations for the distances possess roots that are algebraically multiple, i.e., 
$\Omega$ contains two or more equal valued distances, which leads to configurations 
connected by rotations and mirror reflections. Thus the number of distinct 
configurations associated with the distances is smaller than $k$. For 
$n$-point configurations, the largest $k$ that we have realized is $\hat{k} = n-1$.
Due to space limitation, we could not exhaust all 
degeneracies for each $k$ (i.e., about $C_{120}^k$ cases) in this paper and only provide 
a few specific examples.

\begin{figure}[bthp]
\begin{center}
$\begin{array}{c@{\hspace{1.25cm}}c}\\
\includegraphics[width=4.5cm,keepaspectratio]{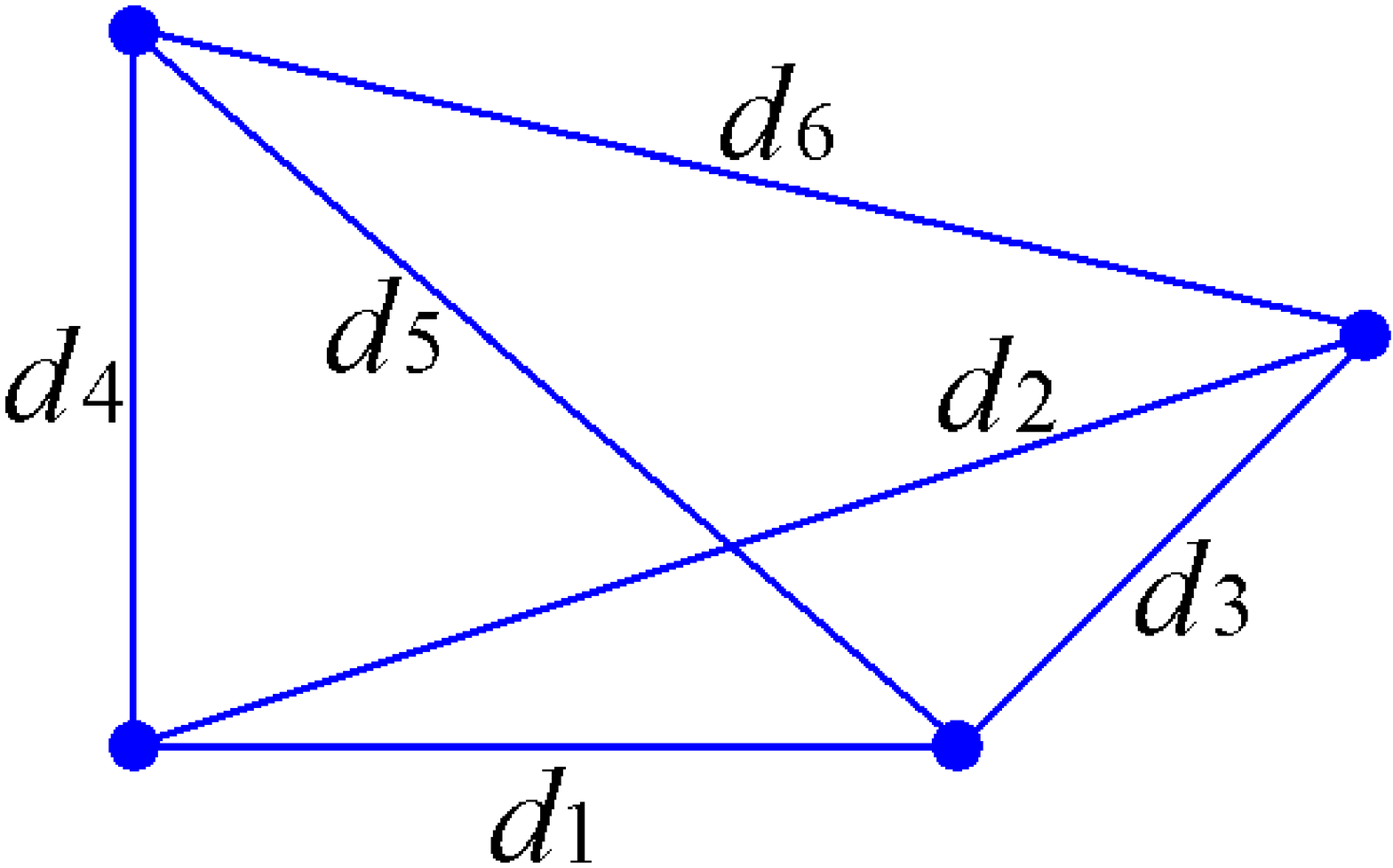} &
\includegraphics[width=4.5cm,keepaspectratio]{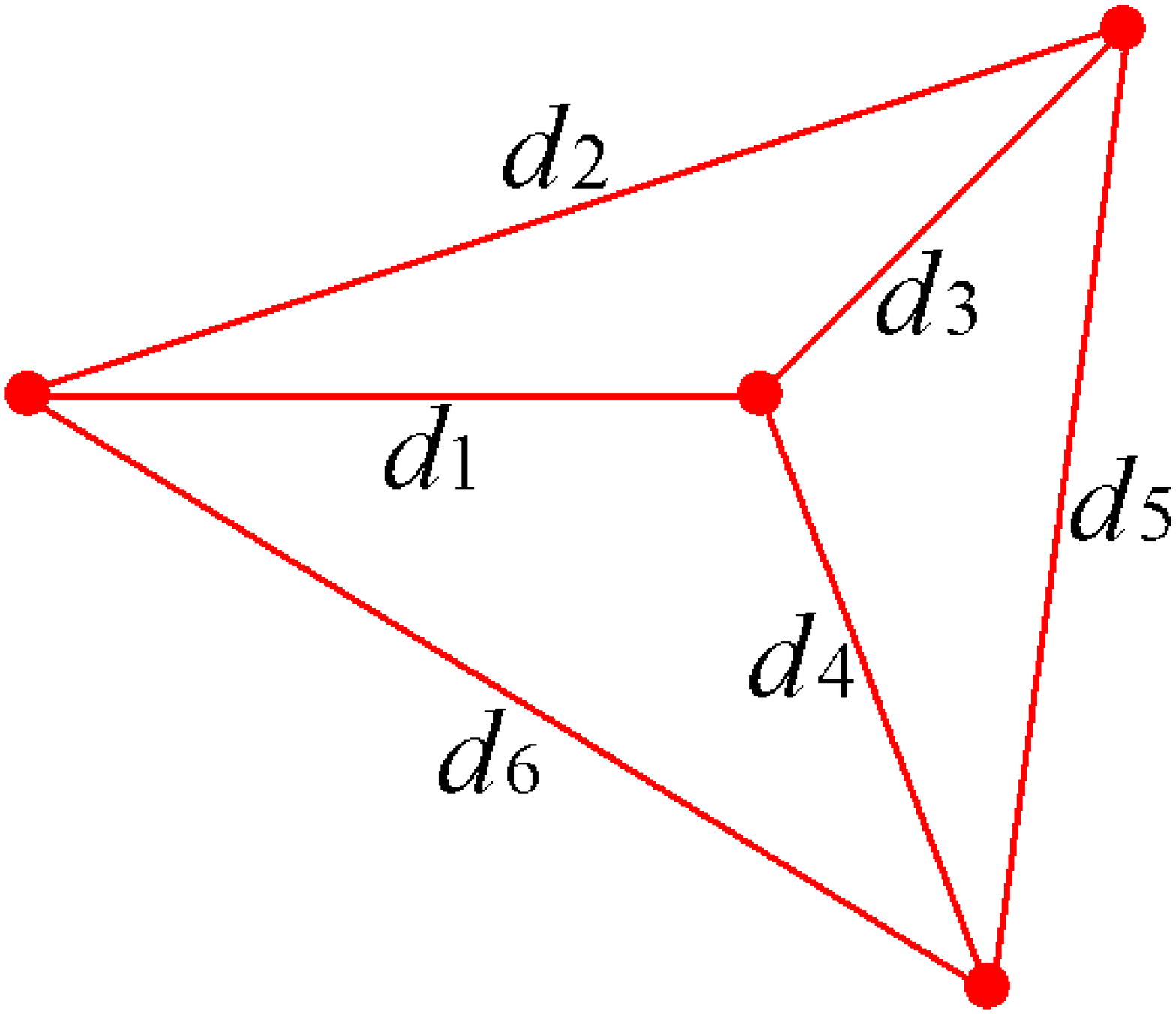} \\
\mbox{\bf (a)} & \mbox{\bf (b)}
\end{array}$
\end{center}
\caption{(color online). An example of two-fold degenerate four-point configurations in $\mathbb{R}^2$. 
The distances are given by $d_1 = 1$, $d_2 = 1.58114\ldots$, $d_3 = 0.70710\ldots$, 
$d_4 = 0.87228\ldots$, $d_5 = 1.32698\ldots$, $d_6 = 1.54551\ldots$.} \label{fig_2fold}
\end{figure}

For $k=2$, requiring the distance permutations $\Omega_1 = \{d_1, d_2, d_3, d_4, d_5, d_6\}$ 
and $\Omega_2 = \{d_1, d_2, d_3, d_6, d_4, d_5\}$ to hold simultaneously yields

\begin{equation}
\label{eq14}
\begin{array}{c}
D(\Omega_1) = D(d_1, d_2, d_3, d_4, d_5, d_6) = 0, \\
D(\Omega_2) = D(d_1, d_2, d_3, d_6, d_4, d_5) = 0,
\end{array}
\end{equation}

\noindent where $D(x_1, x_2, x_3, x_4, x_5, x_6)$ is the multinomial given by 

\begin{equation}
\label{eq15}
D(x_1, \ldots, x_6) = \left|{\begin{array}{c@{\hspace{0.35cm}}c@{\hspace{0.35cm}}c} 
-2x_1^2 & (x_3^2-x_1^2-x_2^2) & (x_5^2-x_1^2-x_4^2)  \\ 
(x_3^2-x_1^2-x_2^2) & -2x_2^2 & (x_6^2-x_2^2-x_4^2)  \\ 
(x_5^2-x_1^2-x_4^2) & (x_6^2-x_2^2-x_4^2) & -2x_5^2 \\\end{array}}\right|.
\end{equation}

\noindent Equation (\ref{eq14}) reduces the free dimensions of the $\mathbb{D}$ space 
from five to four. Without loss of generality, we choose $d_1 = 1$, $d_2 = 1.58114\ldots$, 
$d_3 = 0.70710\ldots$, $d_4 = 0.87228\ldots$, and solve (\ref{eq14}) 
to obtain $d_5 = 1.32698\ldots$, $d_6 = 1.54551\ldots$.
The two-fold degenerate configurations are shown in Fig.~\ref{fig_2fold}.
It should be noted in passing that the ``kite-trapezoid'' example shown 
earlier in Fig.~\ref{kite-trape} is a special case of this four-point 
two-fold degeneracy, for which the shapes each have a reflection symmetry.
If we require the permutations 
 $\Omega_3 = \{d_1, d_2, d_3, d_4, d_6, d_5\}$ 
and $\Omega_4 = \{d_1, d_2, d_3, d_5, d_4, d_6\}$ to hold simultaneously, the 
same degeneracy can be obtained, because the apparently 
different groups of distance permutations $(\Omega_1, \Omega_2)$ and 
$(\Omega_3, \Omega_4)$ correspond to the permutation of indistinguishable points.

\begin{figure}[bthp]
\begin{center}
$\begin{array}{c@{\hspace{0.75cm}}c@{\hspace{0.75cm}}c}\\
\includegraphics[width=4.5cm,keepaspectratio]{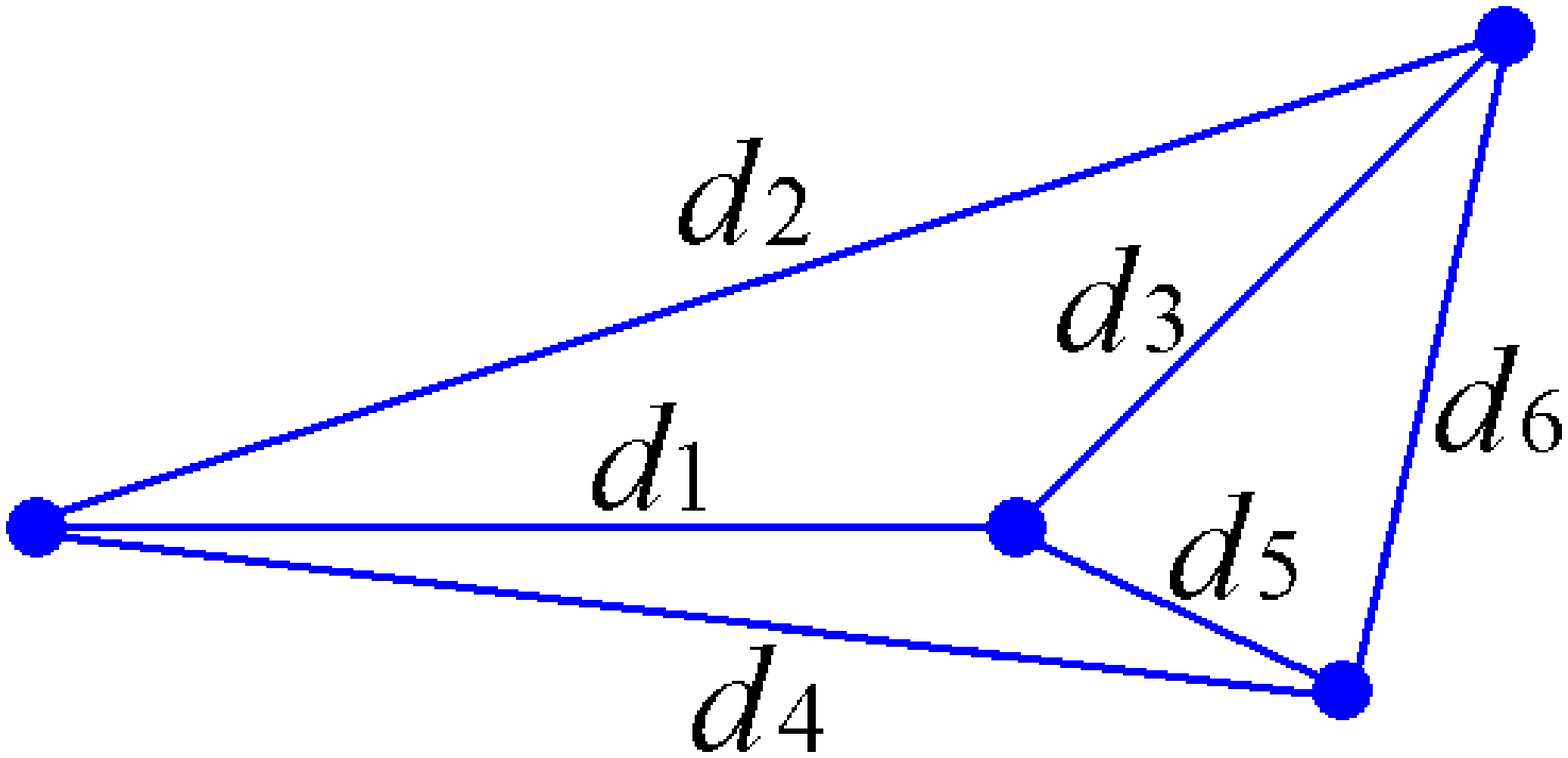} &
\includegraphics[width=4.5cm,keepaspectratio]{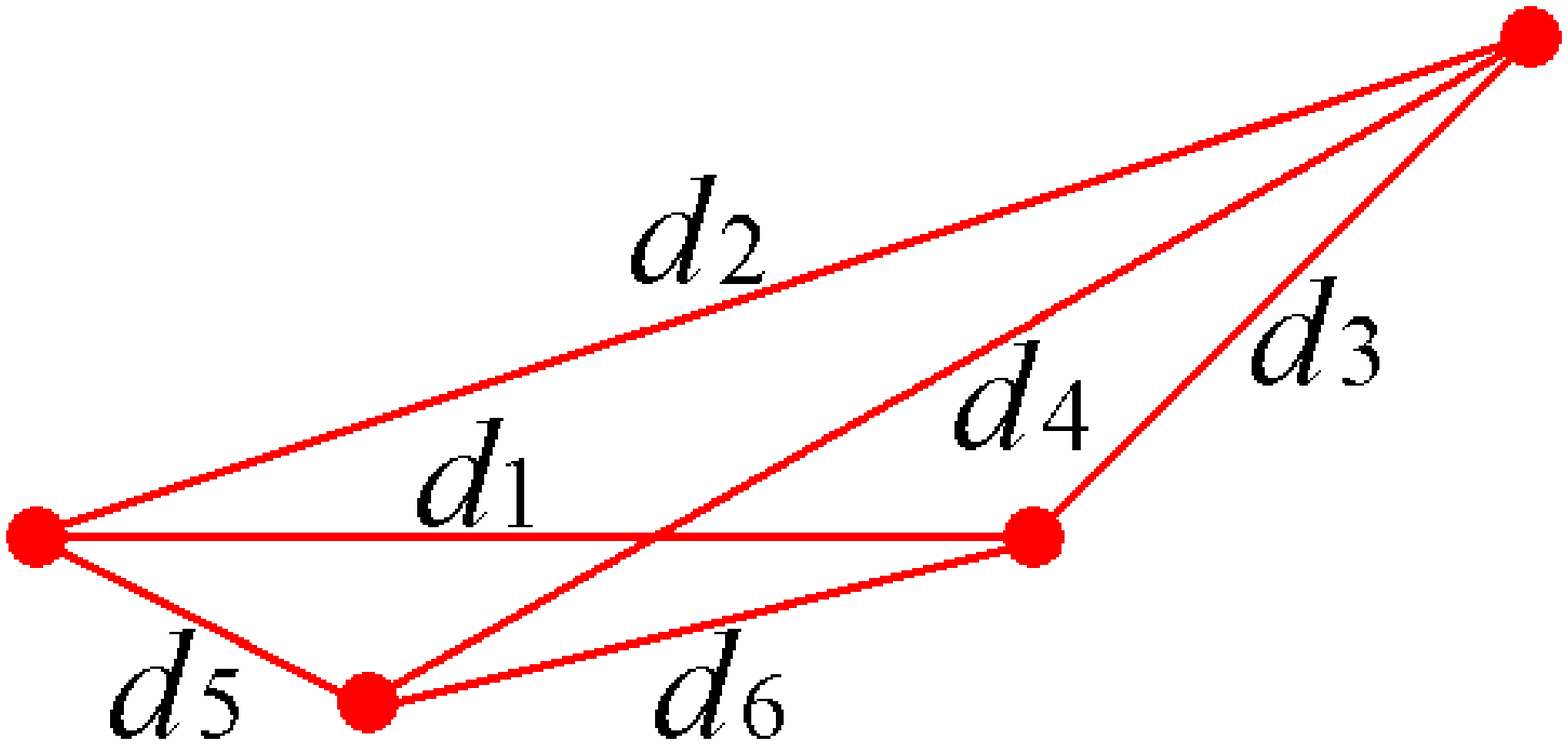} &
\includegraphics[width=4.5cm,keepaspectratio]{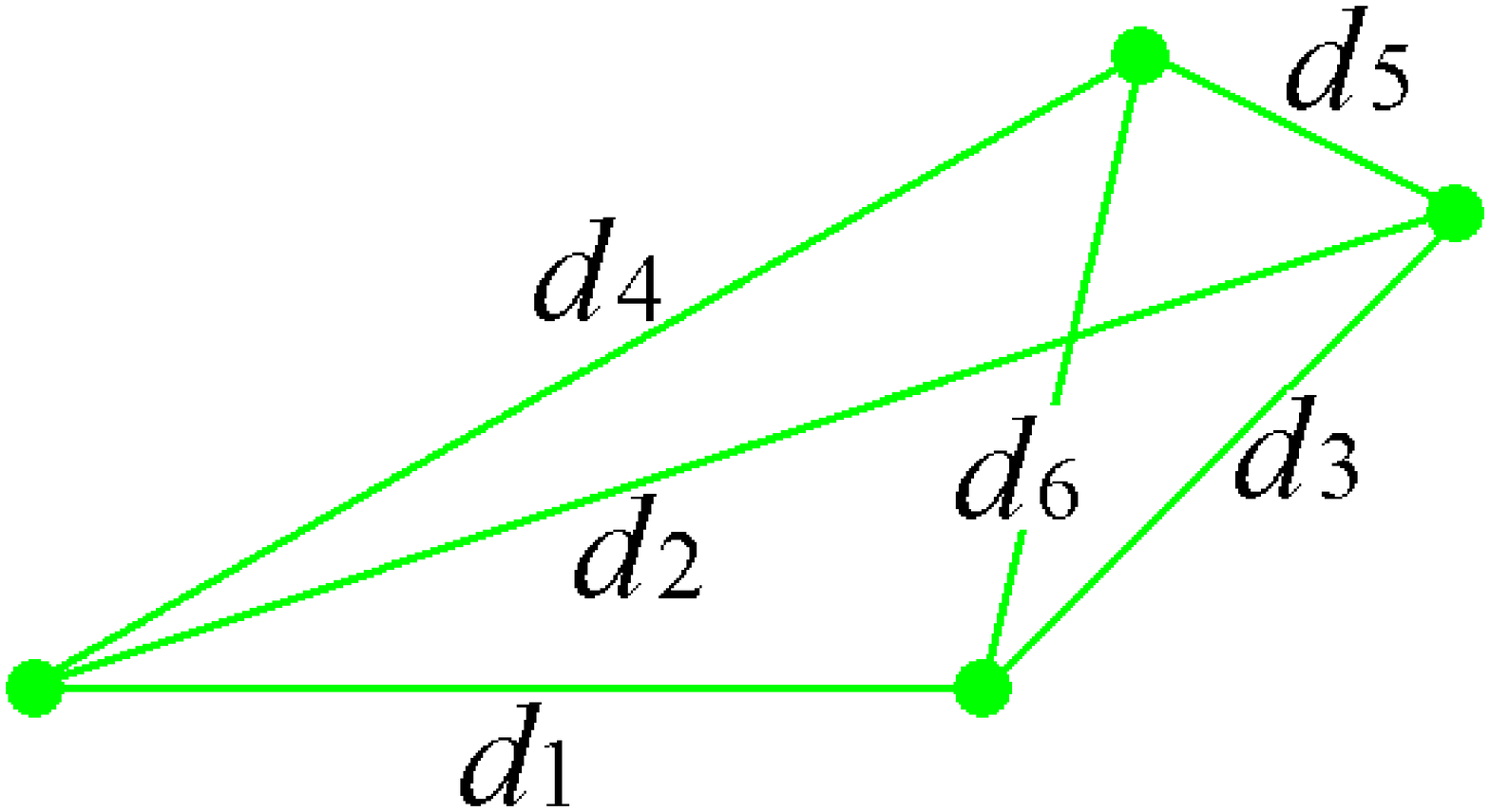} \\
\mbox{\bf (a)} & \mbox{\bf (b)}  & \mbox{\bf (c)}
\end{array}$
\end{center}
\caption{(color online). An example of three-fold degenerate four-point configurations in $\mathbb{R}^2$. 
The distances are given by $d_1 = 1$, $d_2 = 1.581144\ldots$, $d_3 = 0.70710\ldots$, 
$d_4 = 1.34371\ldots$, $d_5 = 0.37267\ldots$, $d_6 = 0.68718\ldots$.} \label{fig_3fold}
\end{figure}

Similarly, for $k=3$ we choose $\Omega_1 = \{d_1, d_2, d_3, d_4, d_5, d_6\}$, 
$\Omega_2 = \{d_1, d_2, d_3, d_5, d_6, d_4\}$ and 
$\Omega_3 = \{d_1, d_2, d_3, d_4, d_6, d_5\}$ to hold simultaneously, 
which reduces the free dimensions to three. By choosing 
$d_1 = 1$, $d_2 = 1.581144\ldots$, $d_3 = 0.70710\ldots$, equations $D(\Omega_i) = 0$ ($i=1, 2, 3$) 
can be solved to yield $d_4 = 1.34371\ldots$, $d_5 = 0.37267\ldots$, $d_6 = 0.68718\ldots$. 
The three-fold degenerate configurations are shown in Fig.~\ref{fig_3fold}.

\subsection{$d$-Dimensional $n$-Point Configurations with Two-Fold Degeneracy}

In general, the feasible region of $k$-fold degenerate $n$-point configurations 
in $\mathbb{R}^d$ can be obtained by carrying out a similar calculation used in the 
previous section, which would be extremely tedious. However, when the point 
configurations possess certain symmetries, particular degeneracies can readily be 
constructed. Here we provide constructions of two-fold degenerate $n$-point 
configurations in $\mathbb{R}^d$ by taking advantage of their symmetries.

\begin{figure}[bthp]
\begin{center}
$\begin{array}{c@{\hspace{0.25cm}}c@{\hspace{0.25cm}}c}\\
\includegraphics[width=4.5cm,keepaspectratio]{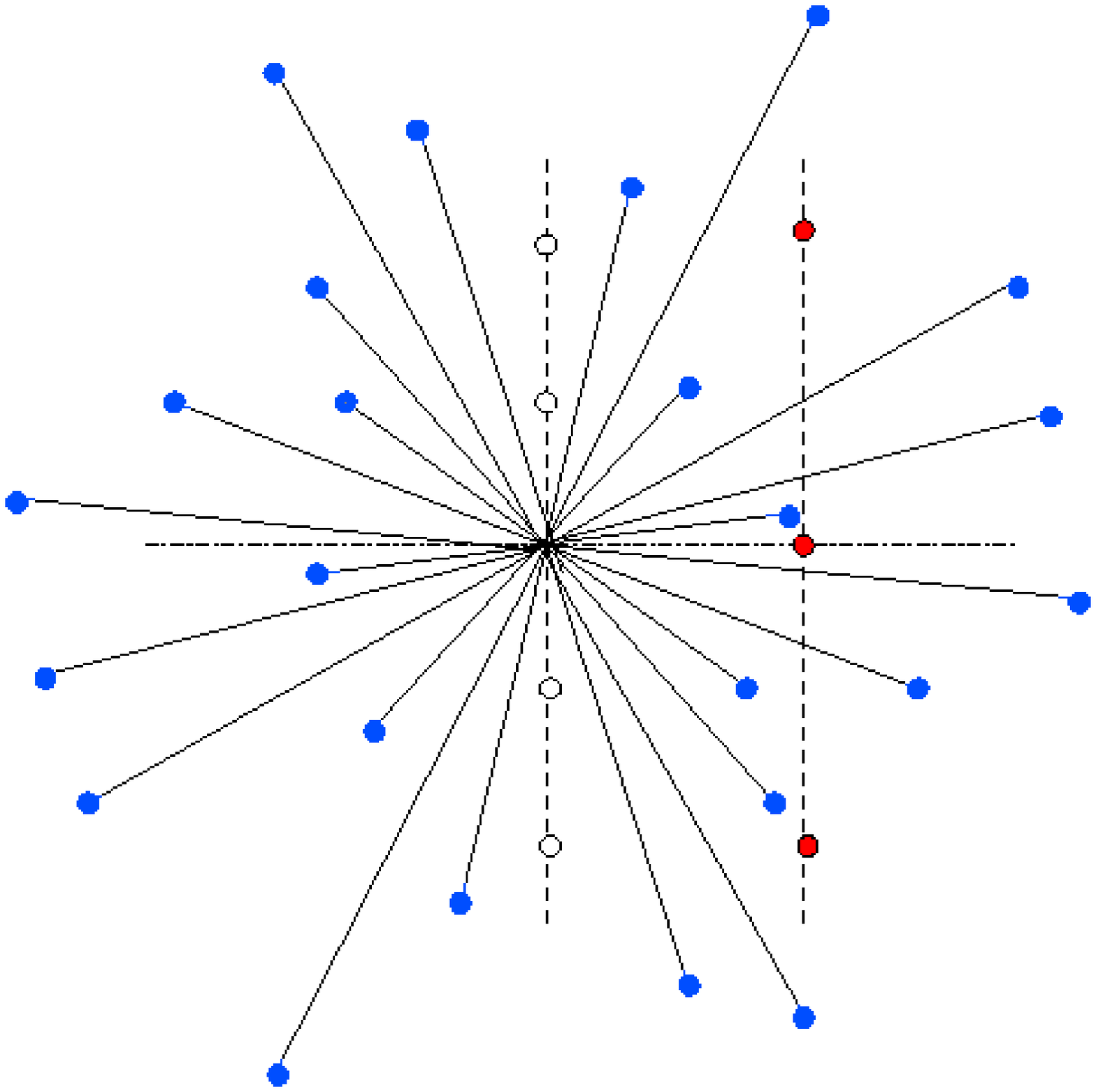} &
\includegraphics[width=4.5cm,keepaspectratio]{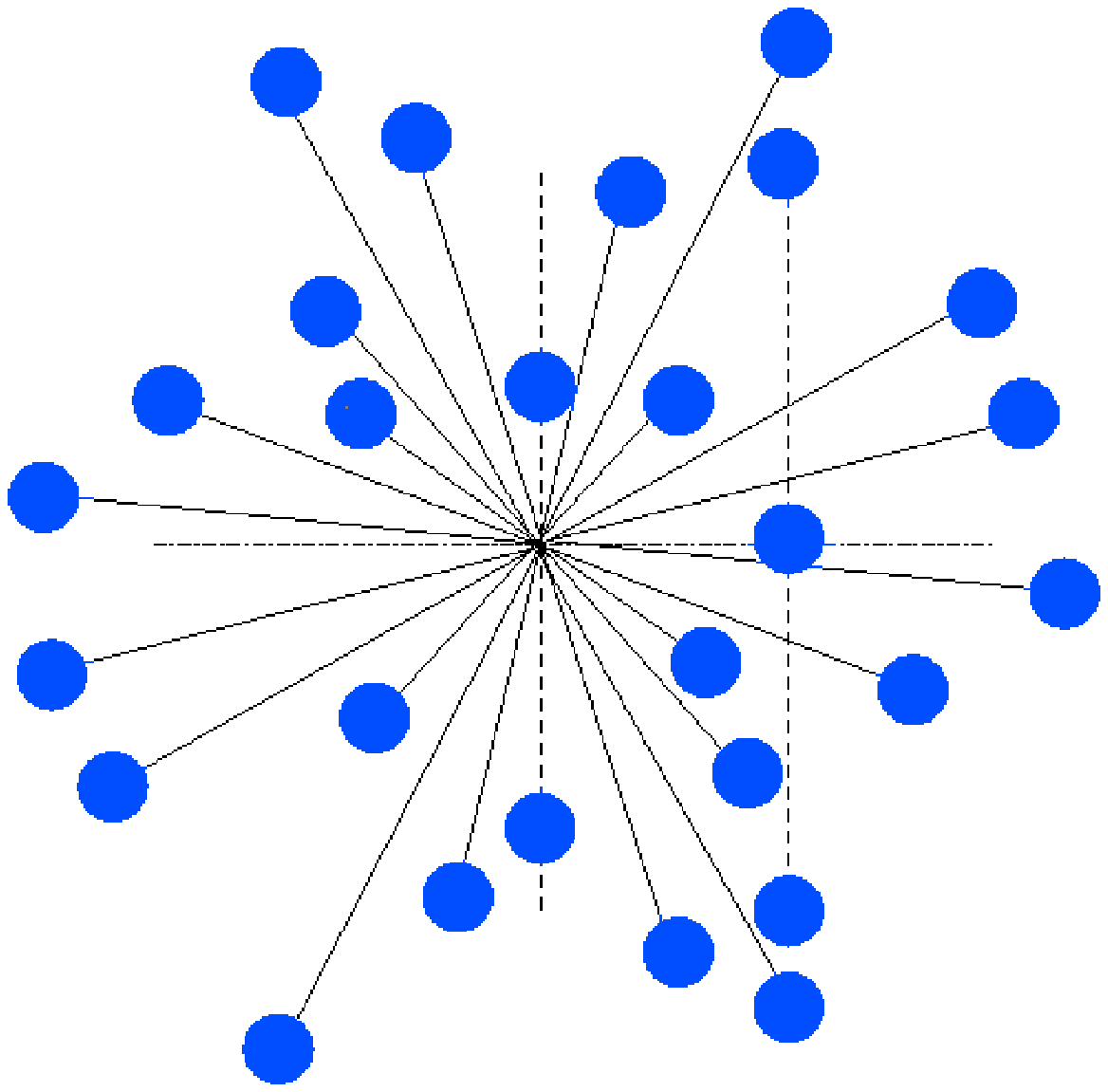} &
\includegraphics[width=4.5cm,keepaspectratio]{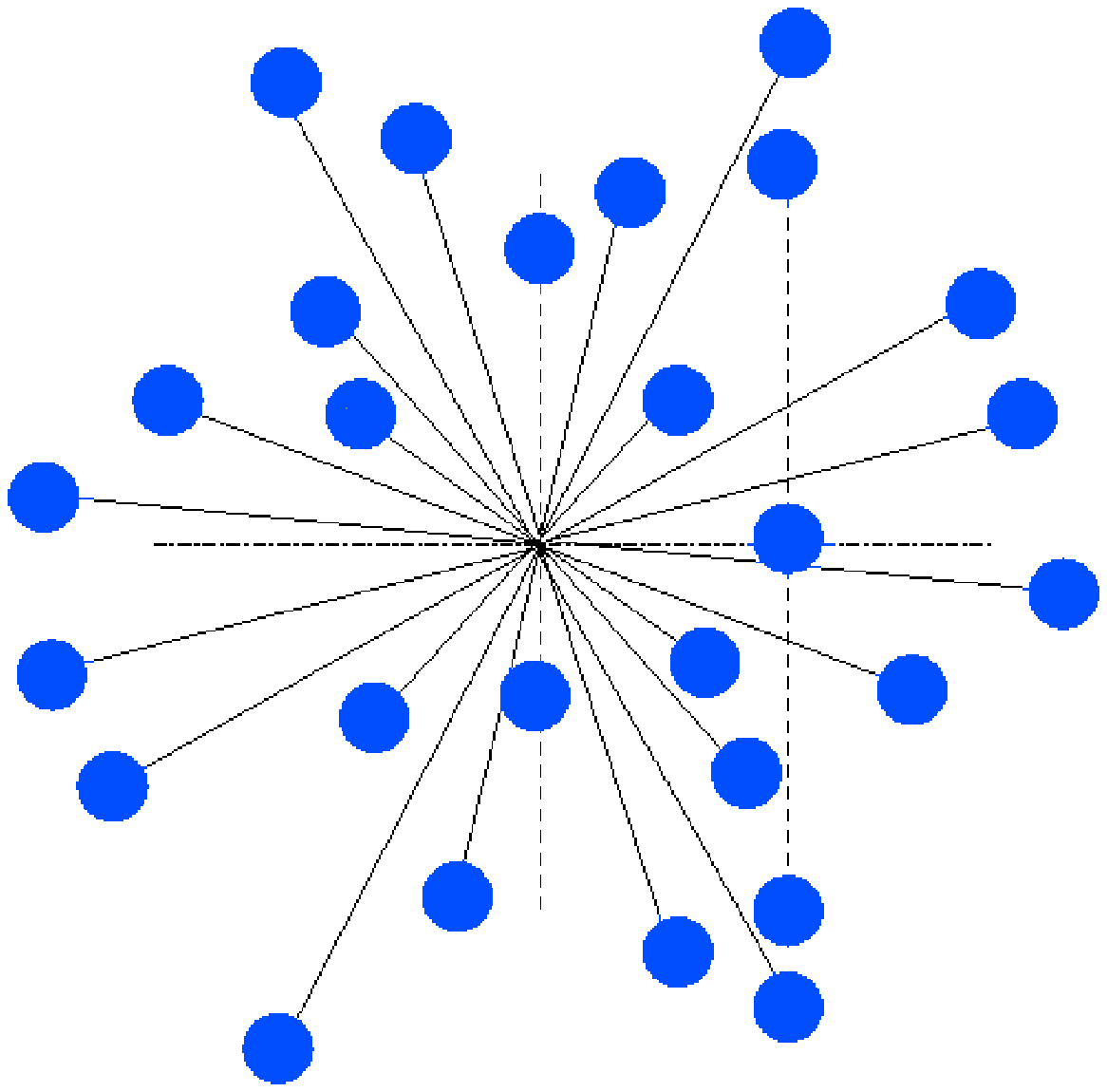} \\
\mbox{\bf (a)} & \mbox{\bf (b)} & \mbox{\bf (c)}
\end{array}$
\end{center}
\caption{(color online). An example of two-fold degenerate configurations in $\mathbb{R}^2$ 
constructed as described in the text. (a) The points in $\Gamma^{(1)}_{d, n_1}$ are shown in 
blue, the points in $\Gamma^{(1)}_{d, n_1}$ are shown in red and the points 
in $\Gamma^{(3)}_{d, 2n_3}$ are shown as void circles. (b) and (c) shows the 
degenerate configuration pair.} \label{fig_2fold2D}
\end{figure}

\begin{figure}[bthp]
\begin{center}
$\begin{array}{c@{\hspace{0.25cm}}c@{\hspace{0.25cm}}c}\\
\includegraphics[width=4.5cm,keepaspectratio]{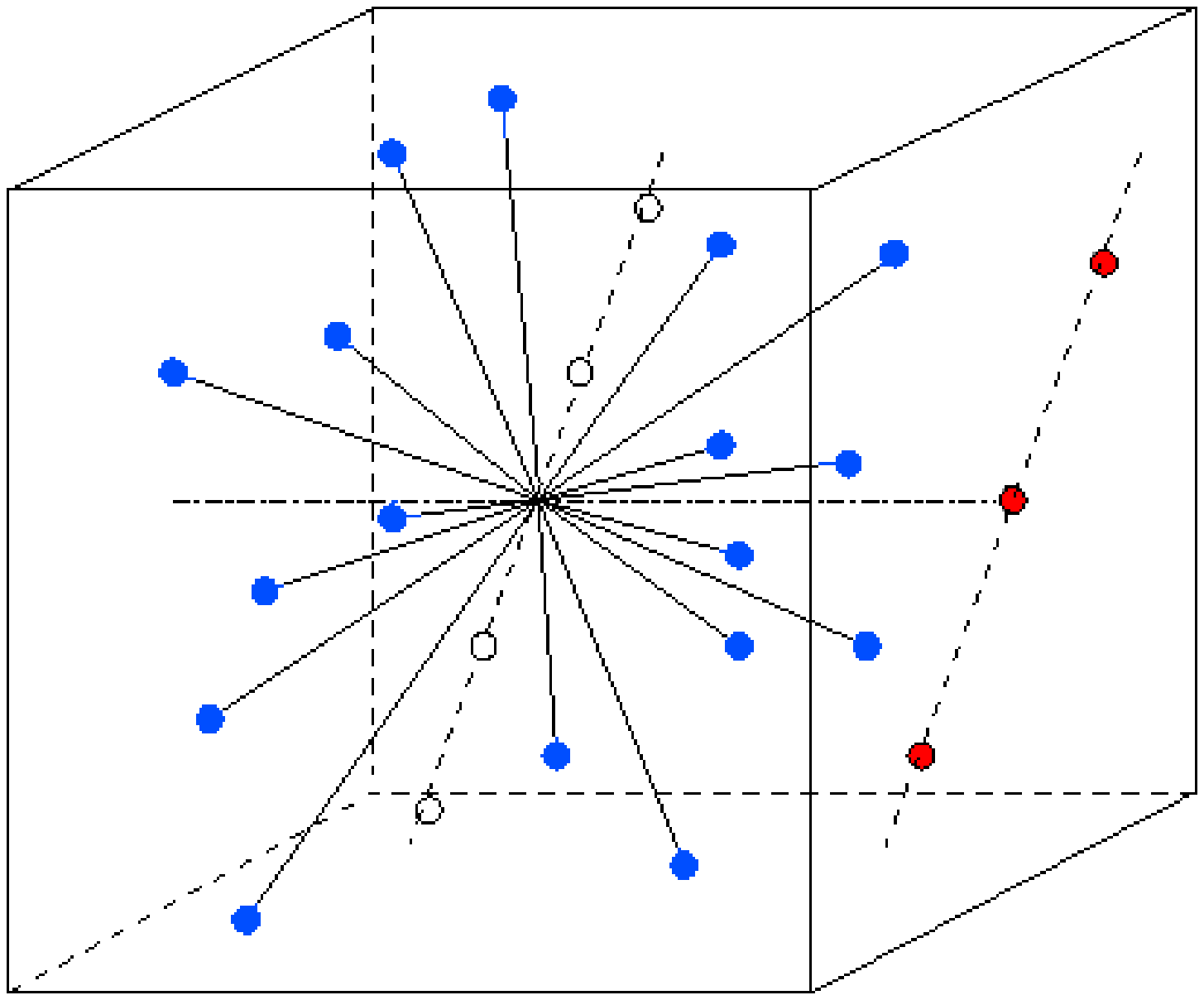} &
\includegraphics[width=4.5cm,keepaspectratio]{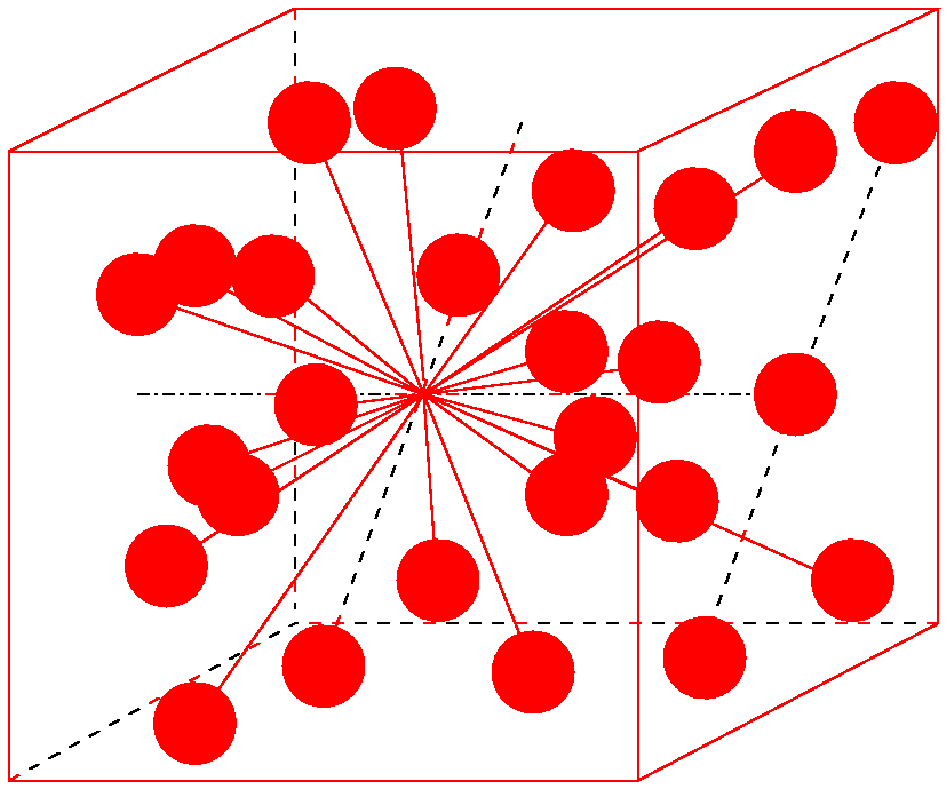} &
\includegraphics[width=4.5cm,keepaspectratio]{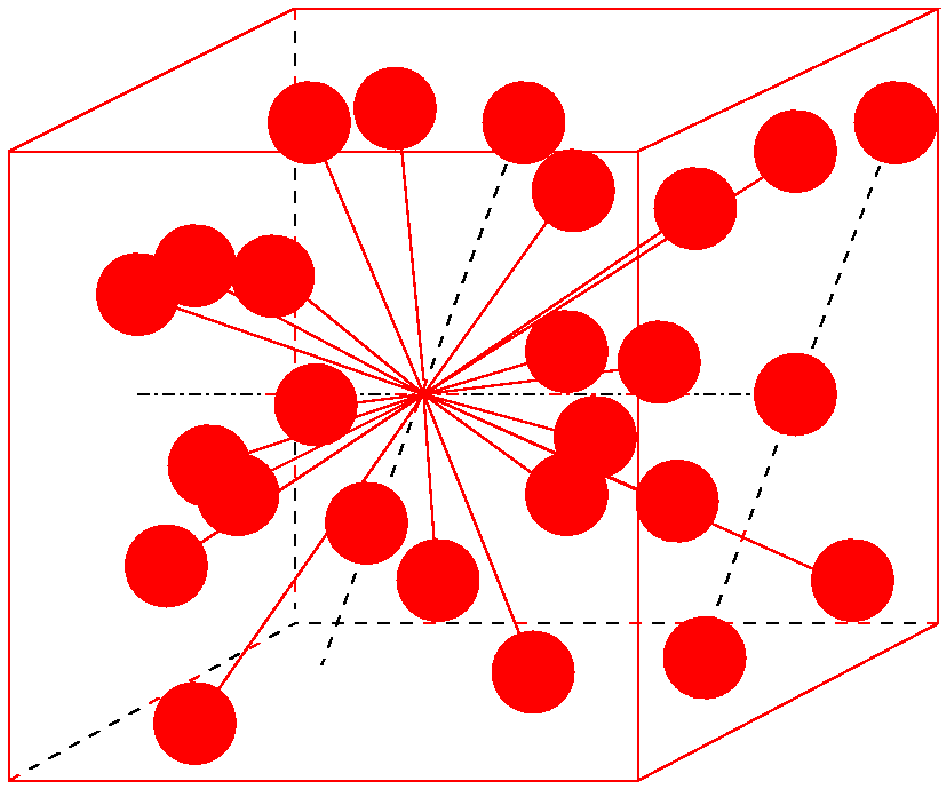} \\
\mbox{\bf (a)} & \mbox{\bf (b)} & \mbox{\bf (c)}
\end{array}$
\end{center}
\caption{(color online). An example of two-fold degenerate configurations in $\mathbb{R}^3$ 
constructed as described in the text. (a) The points in $\Gamma^{(1)}_{d, n_1}$ are shown in 
blue, the points in $\Gamma^{(1)}_{d, n_1}$ are shown in red and the points 
in $\Gamma^{(3)}_{d, 2n_3}$ are shown as void circles. (b) and (c) shows the 
degenerate configuration pair.} \label{fig_2fold3D}
\end{figure}

Consider a \textit{centrally symmetric} $n_1$-point configuration $\Gamma^{(1)}_{d, n_1}$ 
in $\mathbb{R}^d$, i.e., there exists a center $O_1$ such that for every point $P^{(1)}_i$ 
in $\Gamma^{(1)}_{d, n_1}$ there exists a point $P^{(1)}_j$, for which the line segment 
$\overline{P^{(1)}_iP^{(1)}_j}$ passing $O_1$ is bisected by $O_1$ (note that $i=j$ is allowed), 
i.e., $P^{(1)}_i$ and $P^{(1)}_j$ are points of inversion symmetry about $O_1$.  
Consider another centrally symmetric point configuration $\Gamma^{(2)}_{d, n_2}$, 
in which all the $n_2$ points are distributed symmetrically 
on a one-dimensional line $l^{(2)}$ embedded 
in $\mathbb{R}^d$. Denote the symmetry center of $\Gamma^{(2)}_{d, n_2}$ by $O_2$. 
We require that the line segment $\overline{O_1O_2}$ is perpendicular to $l^{(2)}$. 
Finally, consider the centrally symmetric point configuration $\Gamma^{(3)}_{d, 2n_3}$, 
the $2n_3$ points of which are also distributed symmetrically on a one-dimensional line $l^{(3)}$ 
that is parallel to $l^{(2)}$ with the symmetry center coinciding with $O_1$. 
$\Gamma^{(3)}_{d, 2n_3}$ can be further 
partitioned into two subsets: $\Lambda^p_{d, n_3}$ which contains $n_3$ points in 
$\Gamma^{(3)}_{d, 2n_3}$ such that no two points in $\Lambda^p_{d, n_3}$ are symmetric 
about $O_1$ (i.e., they are ``primary'' points); and $\Lambda^d_{d, n_3}$ 
which contains the remaining $n_3$ points of $\Gamma^{(3)}_{d, 2n_3}$ (i.e., 
the ``dual'' points). It is clear that 

\begin{equation}
\label{eq16}
\begin{array}{c}
\Gamma^p_{d,(n_1+n_2+n_3)} = \Gamma^{(1)}_{d, n_1}\cup \Gamma^{(2)}_{d, n_2} \cup \Lambda^p_{d, n_3}, \\
\Gamma^d_{d,(n_1+n_2+n_3)} = \Gamma^{(1)}_{d, n_1}\cup \Gamma^{(2)}_{d, n_2} \cup \Lambda^d_{d, n_3},
\end{array}
\end{equation}

\noindent form a degenerate pair, i.e., the distances from the $n_3$ primary 
points to the remaining $(n_1+n_2)$ points in $\Gamma^p_{d,(n_1+n_2+n_3)}$ are identical 
to those from the $n_3$ dual points to the remaining $(n_1+n_2)$ points in 
$\Gamma^d_{d,(n_1+n_2+n_3)}$, while the two resulting configurations are not 
connected by translation, rotation, mirror reflection or any of their combinations. 
Specific two-fold degeneracy examples in $\mathbb{R}^2$ and $\mathbb{R}^3$ 
are shown in Figs.~\ref{fig_2fold2D} and \ref{fig_2fold3D}, respectively.

\section{Discussion}


The $\mathbb{D}$ space concept can be applied to reconcile 
a variety of problems in statistical physics, such as 
the reconstruction of atomic structures from experimentally obtained $g_2$,
and the decorrelation principle, which we will discuss in 
the ensuing subsections.


\subsection{Reconstruction of Atomic Structures from Experimentally Obtained $g_2$}


As mentioned in Sec. I, a knowledge of atomic structures of condensed matter can 
be obtained via X-ray scattering experiments. In particular, the two-body distribution 
function $g_2(r)$ is related to the Fourier transform of the structure factor $S(k)$, which is 
proportional to the scattering intensities (with the atomic structure function removed). 
For ideal crystalline structures (without any thermal agitation of the atomic centers), $g_2$ consists of a series of 
Dirac delta functions at specific distances. For disordered structures (lack of long-range 
order), $g_2$ is generally a continuous damped oscillating function that decays to 
its long-range value very quickly.  Interestingly, it seems that though 
the pair information contained in $g_2$ of the crystalline matter would determine the 
structures to high accuracy, it is not the case for disordered structures.

The reason can be easily seen if we consider the $\mathbb{D}$ space. For 
an ordered point configuration (i.e., a lattice), there are strong dependencies 
among the distances besides those required by the feasibility conditions. 
For example, consider a $d$-dimensional Bravais lattice whose basis vectors are 
${\bf a}_1, {\bf a}_2, \ldots, {\bf a}_d$. The vector connecting any two 
points in the lattice can then be expressed as 

\begin{equation}
\label{eq17}
{\bf d} = n_1 {\bf a}_1+ n_2 {\bf a}_2 + \cdots  + n_d {\bf a}_d,
\end{equation}

\noindent where $n_i$ ($i=1,\ldots, d$) are integers. Thus,
 the distance $d$ between any two lattice points are given by 

\begin{equation}
\label{eq18}
{d} = \sum\limits_{i=1}^n n_i^2 \langle {\bf a}_i,{\bf a}_i \rangle + \sum\limits_{i\neq j}^n n_i n_j \langle{\bf a}_i,{\bf a}_j\rangle,
\end{equation}

\noindent where $\langle,\rangle$ denotes the inner product of two vectors. Note that 
$\langle{\bf a}_i,{\bf a}_j\rangle = \frac{1}{2}(\langle{\bf a}_i,{\bf a}_i\rangle+
\langle{\bf a}_j,{\bf a}_j\rangle-\langle{\bf a}_i-{\bf a}_j, {\bf a}_i-{\bf a}_j\rangle)$. 
Thus, Eq.~(\ref{eq18}) implies that every distance of an ordered 
point configuration can be obtained if the lengths of the basis lattice 
vectors and the distances between the end points of different basis lattice vectors 
are specified. In other words, Eq.~(\ref{eq18}) further reduces the 
free dimensions of the $\mathbb{D}$ space of the ordered point configuration 
in $\mathbb{R}^d$ to $f = d(d+1)/2$. The additional conditions given 
by Eq.~(\ref{eq18}) significantly reduce the number of feasible permutations 
of the distances. A unique feature of 
the distances for lattices is that the basis lattice vectors are associated 
with the smallest distances.
To completely reconstruct the lattice configuration, 
the smallest $d(d+1)/2$ distances are selected to be 
assembled into a simplex in $\mathbb{R}^d$ defined 
by the common origin and the end points of all the lattice vectors, which in turn 
determines the fundamental cell of the lattice. 
In $\mathbb{R}^2$, three feasible distances uniquely determine a 
triangle and thus, the rhombical fundamental cell. 
In $\mathbb{R}^3$, there are maximally 30 ways that the 
6 distances could be assembled into a tetrahedron. However, even for a 
two-fold degeneracy, the number of equality constraints 
introduced by Eq.~(\ref{eq18}) (i.e., for two permutations of the 6 distances, Eq.~(\ref{eq18}) 
should hold simultaneously) is much larger than the free dimensions 
of the system, which generally rules out all non-trivial solutions. 
Indeed, it is known that for $d\le3$, pair distances are sufficient 
to uniquely determine Bravais lattices. However, in high dimensions, degeneracies of 
Bravais lattices can be constructed \cite{conway}.

For a disordered structure, Eq.~(\ref{eq18}) does not hold and the values of distances 
would form a continuous spectrum in the infinite volume limit. We consider the 
idealized case that there are a finite number of well defined distances and try to reconstruct the 
configuration from them. An important point is that no matter how carefully 
experiments might be carried out, there would still be small but finite errors  
associated with the distances, i.e., $d_i = \hat{d}_i+\epsilon_i$, where 
$\hat{d}_i$ denotes the real value of the distance and $\epsilon_i$ denotes 
the error. Thus, instead of a single point in the $\mathbb{D}$ space, the 
distances correspond to a small uncertainty region with same dimensions as the $\mathbb{D}$ space. 
As we have pointed out, the presence of degeneracies corresponds to a 
feasible region with reduced free dimension, and thus has vanishing ``volume'' 
compared with the feasible region free of degeneracies, which leads us 
to the conclusion that degeneracies are rare in general. However, due to the uncertainties  
of the measured distances, we see that the feasible regions now are 
``finite'' in size compared with those free of degeneracies. 
This explains why in the reconstructions it is hard to exactly recover 
the target configurations, i.e., all configurations associated with the
distances corresponding to the points in the feasible region should be 
considered with equal probability for a ``fair'' reconstruction procedure.    

\subsection{Decorrelation Principle}

Recently, Torquato and Stillinger \cite{To06} proposed a decorrelation principle concerning the 
disordered hard-sphere packings in high dimensional Euclidean space $\mathbb{R}^d$. In particular, 
the decorrelation principle states that unconstrained spatial correlations 
vanish for disordered packings as the spatial dimension becomes large. In other words, 
as $d$ increases, the short-ranged order beyond contact that exists in low dimensions must 
diminish.
This principle has been explicitly
observed in a variety of disordered packings in high dimensions \cite{Decor}.

The centroids of the hard spheres completely determine a packing, 
which can be considered as a point configuration in $\mathbb{R}^d$ 
in which there is a minimum value of pair separation distances $D$ 
(i.e., the diameter of the spheres) due to the nonoverlapping condition. The 
decorrelation principle amounts to the following statement concerning the 
$\mathbb{D}$ space of the configuration: the requirement that the distances 
can not be smaller than $D$ does not affect the occurrence frequency of 
distances with values greater than $D$ in very high dimensions. Note the above 
should be true for any disordered packings, including both dilute and 
jammed packings. It is known that in low dimensions, $g_2 = H(r-D)$ can only be 
maintained for packings with densities less than a critical value \cite{realizable} and 
for disordered jammed packings $g_2$ shows strong short-ranged oscillations \cite{Do05}, 
which is the manifestation of local spatial correlations due to the nonoverlapping 
constraint. In other words, for the jammed disordered packings, 
the requirement that a desired number of distances of value $D$ 
must be realized in the configuration strongly constrains 
the possible values of other distances in low dimensions, 
especially those on the same magnitude of $D$. In high dimensions, 
the above requirement becomes 
less significant in determining the local arrangements of points. 
Consider the construction used in deriving the feasibility conditions, 
to completely determine the position of a point in $\mathbb{R}^d$, 
a ``reference'' structure containing at least $d$ points is used. 
The positions of the points in the reference structure can be 
chosen almost freely subject to the mild constraint that no two points 
can be closer than $D$. As $d$ increases, larger local structures 
(containing more points) can be constructed before the constraints on 
the separation distances between the points begin to play an 
important role. In additions, there are $(d-m)$ ways to arrange 
a point that has fixed distances to $m$ ($m<d$) points in $\mathbb{R}^d$. 
Thus, as $d\rightarrow \infty$, the constraints on the pair-distance values 
imposed by the requirement that a desired number of distances with value 
$D$ must be realized become insignificant, which 
is consistent with the decorrelation principle.

\subsection{Additional Structural Information}

As we have shown, pair-distance statistics in general is not sufficient 
information to completely determine the point configuration. A natural 
question is what additional information could be used to further reduce 
the compatible configurations associated with identical radial distribution functions. 
A conventional choice is the three-body correlation function $g_3$ \cite{Ri66, Ha06}, 
which provides information how the pair distances should be linked 
into triangles.  Though in certain circumstances $g_3$ could provide additional 
information on the point configuration, 
its determination requires additional effort to obtain either theoretically or computationally. 

It has been suggested in Ref.~\cite{Ji09} that instead of incorporating
information contained higher-order versions of $g_2$, namely, $g_3$, $g_4$, etc.,
one might be better served to seek  other descriptors 
at the two-point level, which can be both manageably measured 
and yet reflect nontrivial higher-order structural information. 
One such quantity is the pair-connectedness function $P_2$ \cite{torquato}, 
(i.e., the connectedness contribution to $g_2$), which contains 
non-trivial topological connectedness information of the point configuration.
Note the ``connectedness" in a point configuration can be 
defined in many ways, e.g., one could circumscribe spheres sround each of the points 
and then define that two points are connected if the two associated spheres 
are either contacting or overlapping, for example. Connectedness information contained 
in $P_2$ is distinct from the ``triangular'' information embodied in $g_3$, e.g., 
$P_2$ is sensitive to clustering effects, whereas $g_3$ is  not.

\subsection{Generalization to Two-Phase Media}

\begin{figure}[bthp]
\begin{center}
$\begin{array}{c@{\hspace{1.5cm}}c}\\
\includegraphics[width=4.5cm,keepaspectratio]{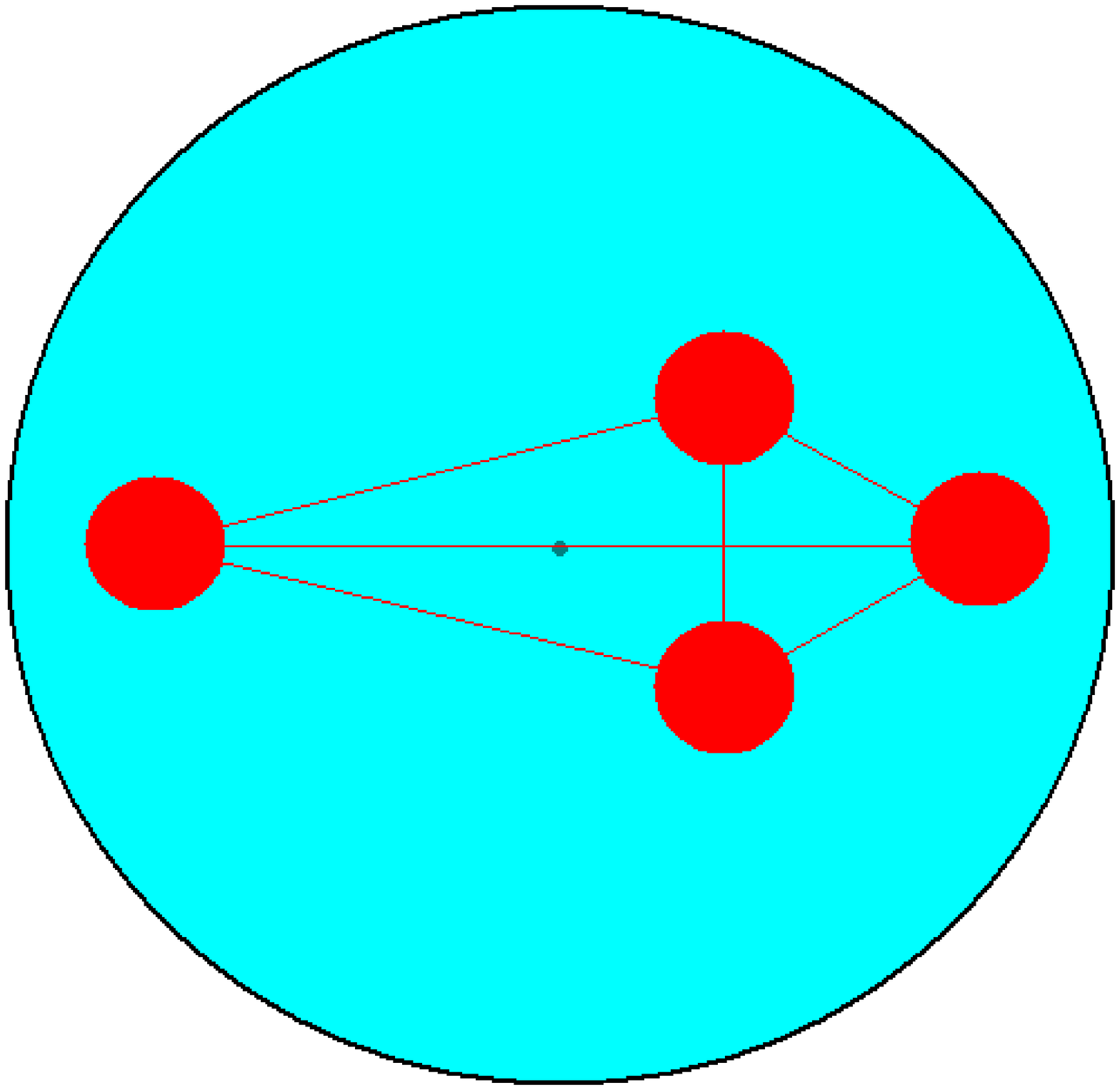} &
\includegraphics[width=4.5cm,keepaspectratio]{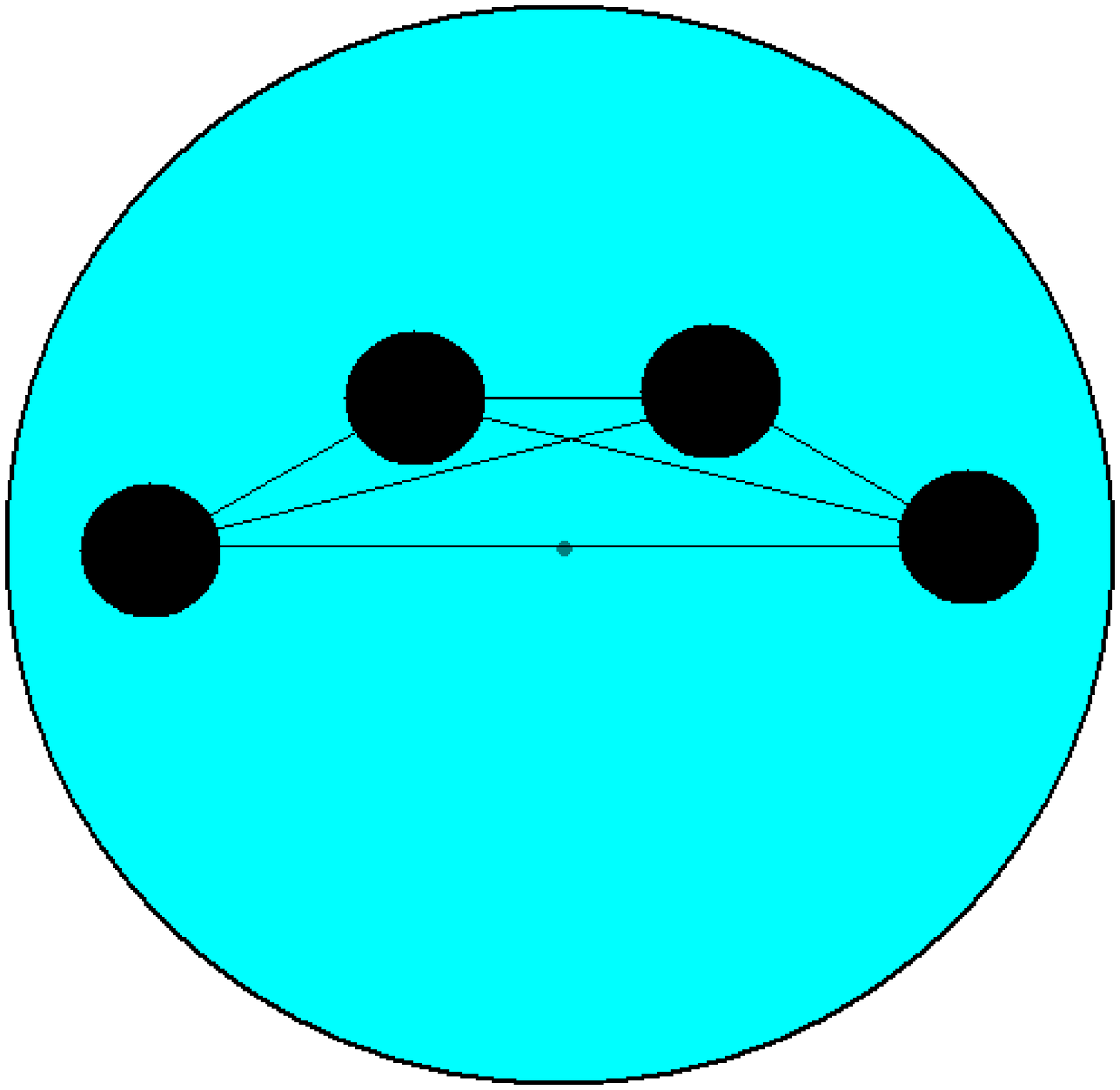} \\
\mbox{\bf (a)} & \mbox{\bf (b)}
\end{array}$
\end{center}
\caption{(color online). An example of two-fold degenerate 
continuum media based on the ``kite-trapezoid" example given 
in Sec.~I. The longest distance in the ``kite'' and ``trapezoid'' 
is symmetrically placed on the large circle diameter.} \label{fig_Medium}
\end{figure}

As pointed out in Sec.~I, two-phase media can be constructed by decorating 
point configurations. For example, one can construct sphere packings by assigning 
to each point a sphere centered at the point with 
diameter equal to the minimal distance in the configuration. 
In this sense, two-phase media are more general than point configurations.
The degeneracy of discrete point 
configurations implies the existence of degenerate two-phase media. 
The corresponding pair-distance information for two-phase media is the two-point 
correlation functions $S_2$ \cite{torquato}. The degeneracy of two-phase media and the 
non-uniqueness issue of their reconstruction will be discussed in 
a sequel (Part II). Here we only provide an example of 
a two-fold degenerate two-phase medium constructed from the ``kite-trapezoid" 
example given in Sec.~I.

As shown in Fig.~\ref{fig_Medium}, suppose we have two large 
solid circles in which small circular holes are made. One large 
circle contains the "kite" holes and the other contains 
the "trapezoid" holes. Since initially the two solid large circles 
are characterized by identical infinite distance set and 
the same subset of distances are then removed to make the holes, 
the remaining sets of distances for the two large circles with holes are 
still identical.

\section{Concluding Remarks}
In this paper, we discussed various aspects of the geometrical ambiguity 
of pair distance statistics associated with general point configurations 
in $\mathbb{R}^d$. In particular, we introduced the idea of the 
$\mathbb{D}$ space and derived the feasibility conditions of the 
distances which are equivalent to the realizability conditions of $g_2$ 
and the necessary conditions for degeneracy. We applied the 
conditions to construct explicit examples of degenerate point configurations 
and showed that though degeneracies are rare, one could not exclude their 
existence merely based on numerical reconstruction studies. We 
also applied the $\mathbb{D}$ space to problems in statistical physics, such as 
the reconstruction of atomic structures from experimentally obtained $g_2$,
and the decorrelation principle.

As pointed out in Sec.~IV.C, the degeneracy of point configurations implies 
the existence of degenerate random media and a simple example is provided there. 
In a sequel to this paper \cite{Ji09b}, we will study the structural degeneracy of general random 
media and the non-uniqueness issue in the reconstruction of heterogeneous materials 
\cite{torquato, Sa03, zohdi, Sal00}.

\begin{acknowledgments}
This work was supported by the American Chemical Society Petroleum Research Fund and the
Office of Basic Energy Sciences, U.S. Department of Energy,
under Grant No. DE-FG02-04-ER46108.
\end{acknowledgments}

\end{document}